\documentclass[fleqn,usenatbib]{mnras}

\usepackage{newtxtext,newtxmath}

\usepackage[T1]{fontenc}

\DeclareRobustCommand{\VAN}[3]{#2}
\let\VANthebibliography\thebibliography
\def\thebibliography{\DeclareRobustCommand{\VAN}[3]{##3}\VANthebibliography}

\usepackage{graphicx}	
\usepackage{amsmath}	

\title[Common envelope evolution of eccentric binaries]{Common envelope evolution of eccentric binaries}

\author[Glanz \& Perets]{
Hila Glanz$^{1}$\thanks{E-mail: glanz@tx.technion.ac.il}
and
Hagai B. Perets$^{1}$
\\
$^{1}$Technion - Israel Institute of Technology, Haifa, 3200002, Israel
}

\date{Accepted 2021 August 2. Received 2021 July 29; in original form 2021 May 5}

\pubyear{2021}

\begin{document}
\label{firstpage}
\pagerange{\pageref{firstpage}--\pageref{lastpage}}
\maketitle

\begin{abstract}
Common envelope evolution (CEE) is believed to be an important stage in the evolution of binary/multiple stellar systems. Following this stage, the CE is thought to be ejected, leaving behind a compact binary (or a merger product). Although extensively studied, the CEE process is still little understood, and although most binaries have non-negligible eccentricity, the effect of initial eccentricity on the CEE has been little explored. Moreover, most studies assume a complete circularization of the orbit by the CE onset, while observationally such eccentricities are detected in many post-CE binaries. Here we use smoothed particle hydro-dynamical simulations (SPH) to study the evolution of initially eccentric ($0\le e\le0.95$) CE-systems. We find that initially eccentric binaries only partially circularize. In addition, higher initial eccentricity leads to a higher eccentricity following the end of the inspiral phase, with eccentricities as high as 0.18 in the most eccentric cases, and even higher if the initial peri-center of the orbit is located inside the star (e.g. following a kick into an eccentric orbit, rather than a smooth transition). CEE of more eccentric binaries leads to enhanced dynamical mass-loss of the CE compared with more circular binaries, and depends on the initial closest approach of the binary. We show that our results and the observed eccentricities of post-CE binaries suggest that the typical assumptions of circular orbits following CEE might potentially be revised.
We expect post-CE eccentricities to affect the delay time distributions of various transients such as supernovae, gamma-ray bursts and gravitational-wave sources by up to tens of percents.  
\end{abstract}




\section{Introduction}
\label{sec:Intro}
Binary common envelope is a key process in the evolution of close binaries. During a common envelope evolution (CEE) a (typically) evolved star fills its Roch-lobe ensuing an unstable mass-transfer. The binary then evolves to a CE phase, leading to the inspiral of the companion inside the envelope and eventually,  the ejection of a significant part (if not all) of the CE \citep[see][for a review]{Iva+13}. The inspiral can give rise to mergers of the companion with the core, or leave behind a compact short-period binary. CEE is believed to play a key role in the evolution of compact systems; the production of X-ray sources and of gravitational wave sources and supernovae progenitors \citep[e.g.][and references therein]{Pac76, Izz+12, Iva+13, Sok17}. 

Due to its relatively short duration and low luminosity, direct detection of systems experiencing this process is almost impossible, with only one candidate system to be observed as an active CE event\citep{2011StepieV1309Evolution,Tyl+11}. Theoretical studies that carry hydro-dynamical simulations of CEE typically fail to fully eject the envelope, and/or to reproduce the orbital properties of post-CE binaries. Various additional physical processes have been suggested to play a role in the CEE and the envelope ejection, such as recombination energy \citealt{Iva+15}, accretion and jets \citep{2019shiber-jetsinCE, 2019schreier-jectsinaTCE}, long pulsations \citep{Cla+17} and dust driven winds \citep{DustDrivenWindsPaper}. The role and importance of each of these various components are still debated. 
Some of these processes (or their combinations) can potentially be distinguished by their timescales \citep[e.g.][]{Michely+19,2019igoshev} and the specific stage of the CE in which they have their strongest influence. The characteristics of the observed late stage outcome system (such as orbital separation and eccentricity) could then potentially identify which of the proposed processes could indeed have an important role in the evolution of the CE. 

In particular, the physical processes and their related timescales and mass ejection could determine the final outcome of the surviving binaries, their period and eccentricity. If most of the mass loss happens in a dynamical time scale, we should expect to observe many post-CE systems with high eccentricities,  whereas a low mass loss rate during the plunge in should lead to a fast circularization of the orbit in most cases, even-though an eccentricity may be developed later-on due to fast episodic mass loss \citep{soker-eccentricitylossinagb,Cla+17}.

Observational studies of short-period binary systems with a sdB component (which might be the result of a RG that lost its envelope during the CE) tend to assume zero eccentricity; although some show non-negligible eccentricities of up to 0.15 \citep{1999Delfosse-eccentric-short-Mdward-WD, 2005Edelmann-eccentric-short-sdBs, 2015Kawka-closesdBseccentric,2021KruckowPotentialPCE}. Moreover, eccentric short-period binary systems with an Asymptotic Giant Branch (AGB) component do exist \citep{2008A&A-formerAGBeccentric}, as well as eccentric long- period binary systems that contain sdB stars \citep{2018MNRAS-sdBSampleEccentric-vos}.

Although CEE have been extensive studied numerically, the vast majority have initiated the binary system with a circular orbit. One reason to believe the orbit has already circularized prior to the onset of the CE, is due to tidal dissipation \citep{1977A&Azahn, 1995verbunt,soker-eccentricitylossinagb}. Nevertheless, as was recently studied by \citet{2020Vigna-Gomez-CEEtoDoubleNS, vick2020tidal}, Roche-lobe overflow of an eccentric binary is not necessarily a special scenario. As described in \citet{soker-eccentricitylossinagb, 2018MNRASKashiTidalCircularizingGrazingEnvelope}, the eccentricity grows dramatically during each pericenter passage due to large mass loss; if the mass loss rate is high enough to overcome the circularization due to dynamical tides, the eccentricity may even increase. 
\\
Eccentric binary systems can also be formed following velocity kicks imparted to neutron-stars and possibly black-holes upon formation in binary systems, or through violent dynamical interactions in destabilized triple systems \citep{per+12c,Mic+14,glanz2020triple} or through secular Lidov-kozai \citep{1962Lidov, 1962Kozai} or quasi-secular \citep{Ant+12} evolution in triple stellar systems \citep[e.g.][and references therein]{2009peretsBlueStragglers,per+12c,Sha+13,Mic+14, toonen2016evolution,Nao16, Moe+17}. These can lead to shorter period binary systems, that can even experience the unstable Roche lobe overflow, leading to the CEE, with very small pericenters, even inside the expanding envelope.

In this paper we study the evolution of eccentric binary systems which experience a CE phase. Here we only follow a gravitational and hydro-dynamical evolution, without considering further processes which might have an important role in the evolution. We first describe our methods and tests, in section \ref{Results} we present our findings; in \ref{sec:Discussion} we further discuss the implications of our results and summarize in \ref{sec:Summary}.

\section{Methods}
\label{sec:Methods}
\begin{figure}
    \centering
    \includegraphics[width=\linewidth,clip]{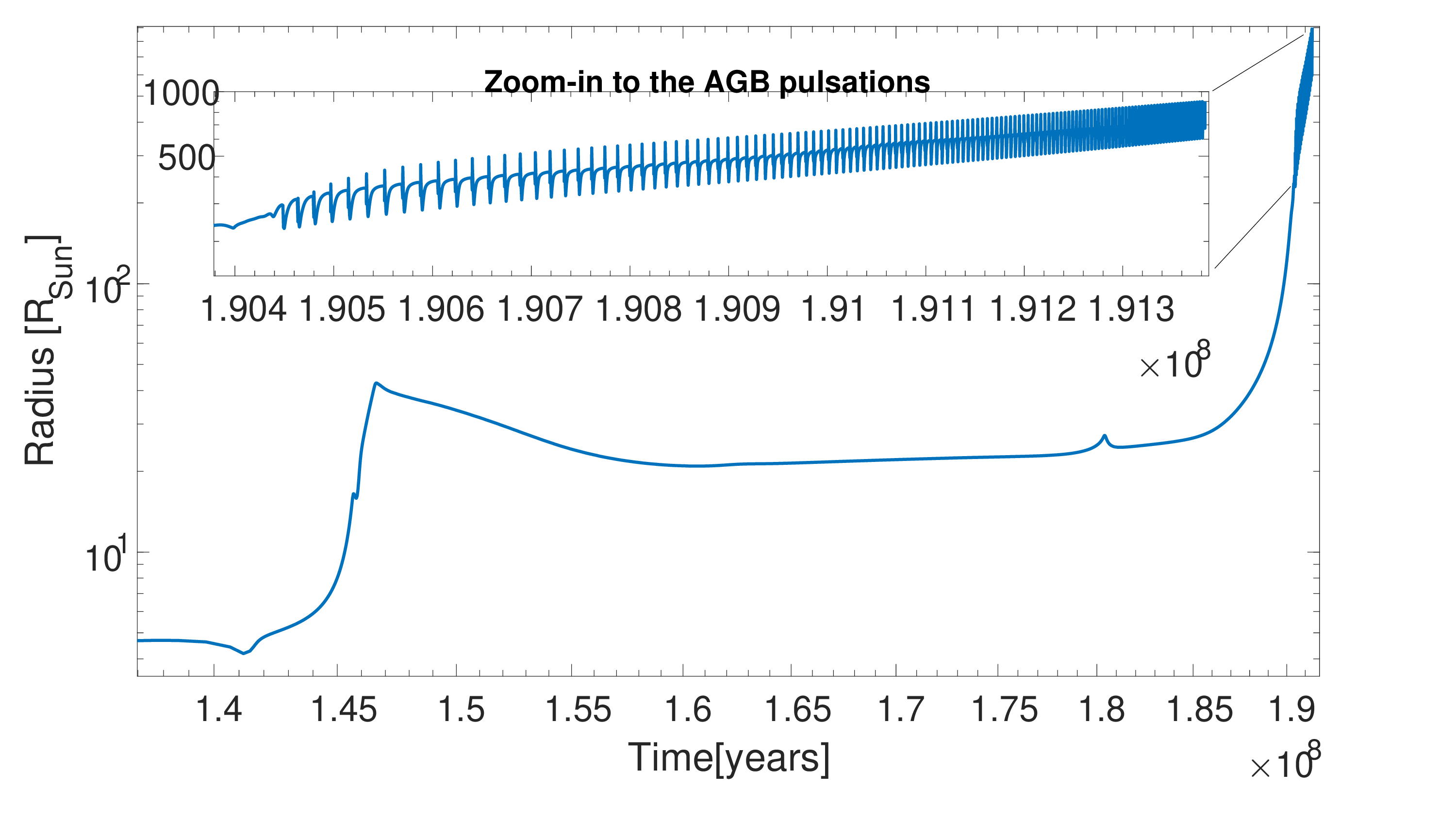}
    \caption{The evolution of the radius of our $4M_\odot$ model, simulated with \texttt{MESA} \citep{mesa-article}}
    \label{fig:4evolution}
\end{figure}

We simulated the common envelope phase of systems with different mass ratios and different initial eccentricities. As in \cite{glanz2020triple}, we used the \texttt{AMUSE} framework \citep{amusearticle} to integrate between the different codes and to analyze the results. We used \texttt{MESA} stellar evolution code (version 2208, \citealt{mesa-article}) to create models of $1M_{\odot}$, $4M_{\odot}$ and $8M_{\odot}$ that were evolved from zero age main sequence into red giants, all have convective envelopes, with the corresponding core masses- $0.39M_\odot$,$0.48M_\odot$ and $1.03M_\odot$ (detailed properties are given in tables \ref{configurations-peri}, \ref{configurations-apo}, \ref{results-table}). While the stages of our $1M\odot$ and $8M_\odot$ were chosen as in \citet{Pass+12} and \citet{ glanz2020triple}, the model with $4M\odot$ was evolved to the tip of its Red giant phase, where it remains for much longer than the dynamical timescale expected for the CE phase (see Fig. \ref{fig:4evolution}). We then converted the 1-dimensional models into 3-dimensional Smoothed Particles Hydro-dynamical (SPH) models with 250K particles (see \citealt{glanz2020triple} for more details about the mapping). As we are not interested in resolving the evolution of the internal structure of the giant core, both the core and (later-on) the companion are modeled as gravitational only interacting particles (also termed dark matter particles in \citealt{gadget2}, due to their use in cosmological simulations). For each giant model, we chose the softening lengths of the point mass particles as in \citet{glanz2020triple}, as long as those satisfy our stability tests as described in Subsection \ref{stability}.

We used the SPH code \texttt{GADGET2} \citep{gadget2} to simulate a relaxation phase and finally the common envelope phase. Within \texttt{GADGET2}, we use the ideal gas EOS for the gaseous particles, including their internal (thermal) energy and gravitational potential, and excluding other energy sources such as radiation pressure and recombination. To initialize a stable model and avoid nonphysical perturbation cause by the change of equation of state and dimensional model, one need to operate a relaxation/damping phase after mapping the 1D \texttt{MESA} model into the SPH model. The model first run in isolation for a few dynamical timescales (we chose $130$ days which in all giant models are longer than 20 dynamical times). After each step of this stage, the positions and velocities of the SPH particles are adjusted such that its center of mass do not change, and the velocities are damped in a decreasing power as follows: \[
r^i_{j}=\left(r^i_{j}-r^i_{COM}\right)+r^0_{COM}
\]
\[
v^i_{j}=\left(v^i_{j}-v^i_{COM}\right)\cdot\left(i/nsteps\right)+v^0_{COM},
\]
Where $r^i_{j}$ and $v^i_{j}$ are the j'th gas particle position and velocity at step $i$, $r^i_{COM}$ and $v^i_{COM}$ are the giant's center of mass position and velocity at step i, $nsteps$ is the total number of the damping steps. After each step the internal velocities are damped by multiplying with a factor which increases from 0 in the first step to 1 in the last one.
Finally, we simulated the common envelope (CE) phase with various simulation times, up to 2850 days, and terminated the simulations in case of a core merger (see \citealt{glanz2020triple} regarding the definition of a merger in our simulations). We note that we only simulate the dynamical phase of the CE, until the orbital parameters do not change for a few orbits. Therefore, the calculated system parameters we present here might be modified later-on by other processes that act in longer timescales. In Sec \ref{massloss} and \ref{timescales} we describe how these parameters can be affected in the latter evolution, and how one can compare our results to the observed values.

\subsection{Common envelope initiation in eccentric binary systems}
\label{initializationInRL}
\begin{figure}
    \centering
    \includegraphics[width=\linewidth,clip]{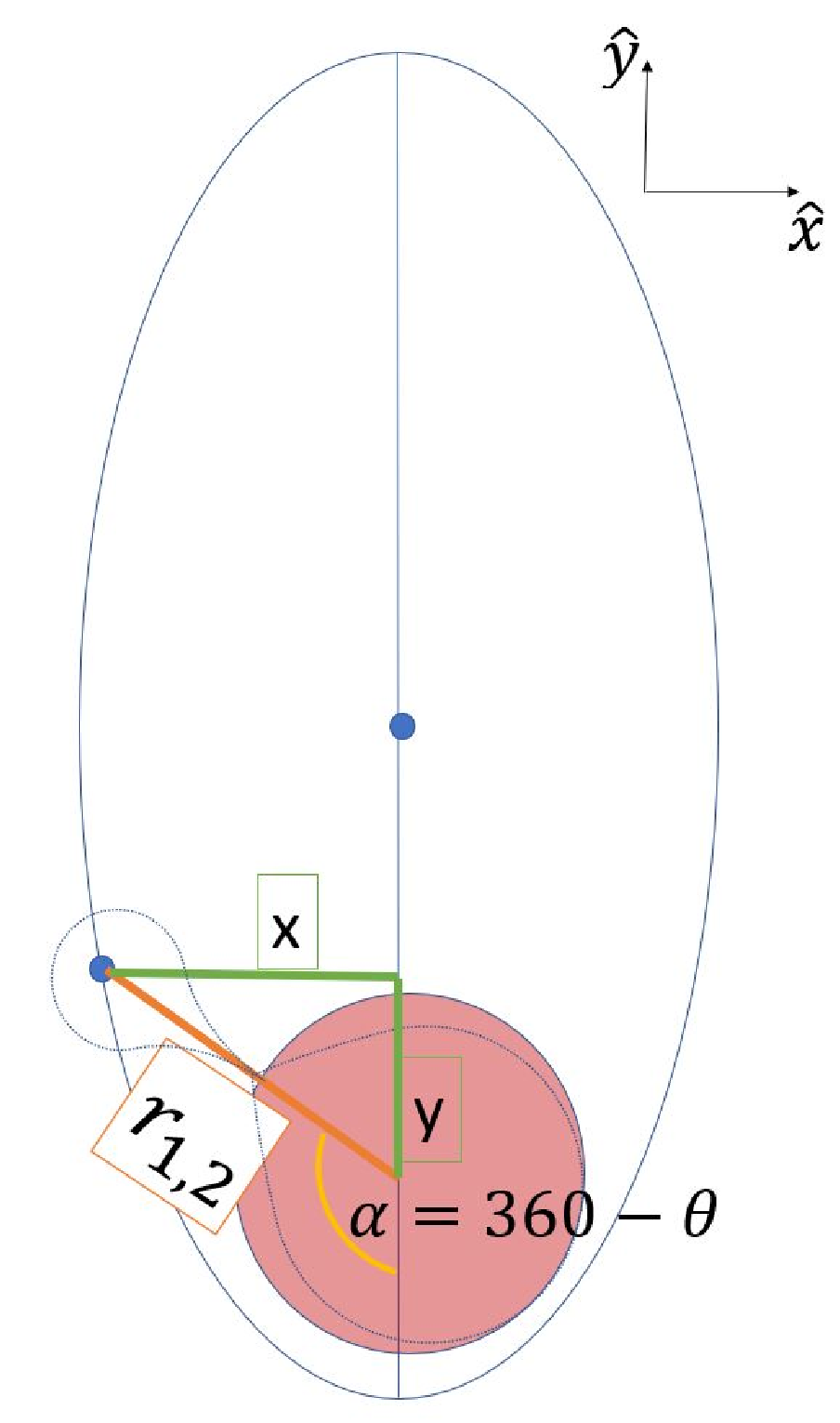}
    \caption{An eccentric binary orbit at the point where the giant fills its Roche-lobe. The y axis is pointing towards the apocenter.}
    \label{fig:beginAtRochelobeFilling}
\end{figure}

Highly eccentric systems can have negligible tidal interaction throughout a large part of their orbits, with an approximately Keplerian motion. The binary separation during a Keplerian orbit with semi-major axis (SMA) $a$ and orbital eccentricity $e$, is:
\begin{equation}
\label{keplerian_r}
    r_{1,2}=\frac{a\left(1-e^2\right)}{1+e\text{cos}(\theta)}
\end{equation}
If we choose the y axis to be the one pointing towards the apocenter (See Fig. \ref{fig:beginAtRochelobeFilling}), the components of the positions and velocities can be written as follows:
\begin{equation}
    x=r_{1,2}\text{sin}\theta \\
    y=-r_{1,2}\text{cos}\theta \\
\end{equation}
\begin{equation}
    v_x=\sqrt{\frac{GM}{a\left(1-e^2\right)}}\left(e+\text{cos}\theta\right) \\
    v_y=\sqrt{\frac{GM}{a\left(1-e^2\right)}}\text{sin}\theta
\end{equation}
$\theta=-\pi$ at the apocenter and 0 at the pericenter.

The common envelope is induced by an unstable Roche lobe overflow of one of the components, most likely by its evolutionary growth. 
The Roche-lobe size of the binary system can be approximated by the following expression by \cite{1983eggleton}:
\begin{equation}
\label{Roche-lobe-radius}
    R_\text{RL}=\frac{0.49q^{2/3}}{0.6q^{2/3}+\text{ln}(1+q^{1/3})}
\end{equation}
Where q is the mass ratio between the primary and secondary. Therefore, the distance between both components where the primary fills its Roche lobe is:
\begin{equation}
\label{overfill_r}
r_{1,2}=R_{1,\text{RL}}=\frac{R_1}{R_\text{RL}}
\end{equation}
Using eq. \ref{keplerian_r} with the proper separation, one can derive the angels at which the orbit fills the giant's Roche-lobe:
\begin{equation}
    \theta_\text{RL}=\text{arccos}\left(\frac{aR_\text{RL}}{eR_1}\left(1-e^2\right)-\frac{1}{e}\right)
\end{equation}
To reduce cumulative numerical errors and to save time, we can begin the simulations of the extreme cases at the point when their orbital separation is more than 2.5 times the one defined in eq. \ref{overfill_r}. More details regrading this method and its verification can be found in Subsection \ref{stability}.

\subsection{Stability tests}
\label{stability}
\subsubsection{Testing the evolution of the SPH model in isolation}
\label{isolationTest}
As some of our models are initiated when the binary components are at separations (apocenter) much larger than the Roche-radius and evolved for long times, before experiencing close-interactions it is necessary to verify the stability in terms of error accumulation of our system in isolation for at least the duration until the beginning of the fast inspiral stage. We used the same method described in \citet{glanz2020triple}, by evolving the system up to $\approx50$ dynamical times and compared the SPH velocities to the typical velocities of the system. We define our stability criteria such that less than 1 percent of the particles gain velocities which are comparable to the typical values (i.e. relative Keplerian velocity of the companion, the sound speed, and the escape velocity). Using our default softening length, we found that for the giant with $1M_\odot$, non-negligible particle velocities were excited after $\approx45$ dynamical times. These timescales are sufficiently long for most of our simulations, however the cases with initial eccentricities of $0.9$ and $0.95$ have longer dynamical timescales and require longer evolution to study the full CEE. Beginning those simulations at a later position on the orbit, when the binary still avoids interactions (as defined in the next Subsec \ref{lowInteractionTests}), in a addition to a  10 times larger core softening length in the most extreme case of $e_i=0.95$, induced much smaller artificial excitation of the velocities which then satisfied our stability criteria.

\subsubsection{Testing the evolution during low interaction parts of the orbit}
\label{lowInteractionTests}
Due to the large part of the orbit in which our binary systems have negligible tidal interactions but might be affected by numerical error accumulating over time, we verified whether an artificial deviation of the orbit from the purely Keplerian orbit occurs. After each pericenter passage, when the separation becomes $2.5R_L$, i.e. the binary components effectively do not interact hydro-dynamically, but only due to gravitational Keplerian motion, we calculate the new Keplerian orbit corresponding to the current system parameters. For each orbit, we calculated the maximal deviation of the orbit in the part between the two points where the separation was about 2.5 times the Roche lobe radius. Since the tidal force goes like $F_\text{Tides}\sim\left(R_1/r_{1,2}\right)^5$ (see \citep{1981hut}), we can write- $r_{1,2}=A\cdot R_1/R_\text{RL}$ and then 
\begin{equation}
F_\text{Tides}= F_\text{Tides}(\theta_\text{RL})\cdot A^{-5}
\end{equation}  At $2.5R_\text{RL}$ distance between the binary components, the configuration is outside the Roche lobe radius, and is far enough such that tidal affects can be neglected. We consider a system to be stable when the maximal deviation from the Keplerian orbit is not more than 10\% of the separation at the point, and is not more than 20 \% of the minimal separation of orbit. We find that the orbit of the most eccentric systems, if initialized at apocenter accumulates too much errors by the time they evolve to pericenter, and we therefore initialize them close to peri-center (but at $>2.5R_\text{RL}$). Following the first pericenter approach these orbits dissipate and the their next apocenter is significantly reduced (see Fig. \ref{fig:095test}). We find that at no time in the later evolution do the error accumulation significantly affects orbits and these systems then pass our criteria. 

\section{Results}
\label{Results}

\begin{table*}
\begin{tabular}{|c|c|c|c|c|c|c|c|c|c|}
\hline 
 Sim & $M_\text{1}$ &$M_\text{2}$ &$M_{1,\text{core}}$ &$R_\text{1}$ &$r_{\text{i}}^{\text{a}}$ &$a_{\text{i}}$& $e_{\text{i}}$ & $R_{1,\text{RL}}$ 
\\ &(${\rm M_{\odot}}$)&(${\rm M_{\odot}}$)&(${\rm M_{\odot}}$)&(${\rm R_{\odot})}$&$({\rm R_{\odot})}$ &$({\rm R_{\odot})}$&&$({\rm R_{\odot})}$\tabularnewline
\hline 
\hline 
1R06-0 &1 & 0.6 & 0.388 & $83$ & 83 & 83 & 0.0 & 34  \tabularnewline
1R06P2 &1 & 0.6 & 0.388 & $83$ & 124.5 & 103.8 & 0.2 & 52  \tabularnewline
1R06P5 &1 & 0.6 & 0.388 & $83$ & 249 & 166 & 0.5 & 103  \tabularnewline
1R06P7 &1 & 0.6 & 0.388 & $83$ & 470.3 & 276.7 & 0.7 & 195  \tabularnewline
1R06P9 &1 & 0.6 & 0.388 & $83$ & 1577 & 830 & 0.9 & 668  \tabularnewline
1R06P95 &1 & 0.6 & 0.392 & $83$ & 3237 & 1660 & 0.95 & 1372  \tabularnewline

\hline
8R2G-0 & 8 & 2 & 1.03 & 110 & 115 & 115 & 0.0 & 58 \tabularnewline
8R2G2 & 8 & 2 & 1.03 & 110 & 172.5 & 144 & 0.2 & 86 \tabularnewline
8R2G5 & 8 & 2 & 1.03 & 110 & 345 & 230 & 0.5 & 173 \tabularnewline
8R2G7 & 8 & 2 & 1.03 & 110 & 651.7 & 383 & 0.7 & 326 \tabularnewline

\hline 
4R06-0 & 4 & 0.6 & 0.478 & 42 & 43 & 43 & 0.0 & 23.4 \tabularnewline
4R06P2 & 4 & 0.6 & 0.478 & 42 & 64.5 & 54 & 0.2 & 35.1 \tabularnewline
4R06P5 & 4 & 0.6 & 0.478 & 42 & 129 & 86 & 0.5 & 70.3 \tabularnewline
4R06P7 & 4 & 0.6 & 0.478 & 42 & 243.7 & 143.3 & 0.7 & 133  \tabularnewline
4R06P9 & 4 & 0.6 & 0.478 & 42 & 817 & 430 & 0.9 & 445.4  \tabularnewline
4R06P95 & 4 & 0.6 & 0.478 & 42 & 1677 & 860 & 0.95 & 914.2 \tabularnewline
\hline 
\end{tabular}\caption{\label{configurations-peri} Initial configuration of the simulated systems with a pericenter located on the edge of the primary's envelope. $M_\text{1}$ is the mass of the primary at zero age in the main sequence, $M_\text{2}$ is the mass of the secondary, $M_{1,\text{core}}$ and $R_\text{1}$ are the core mass and radius of the primary at the beginning of the common envelope, $r_{\text{i}}^{\text{a}}$ is the initial apoapsis of the orbit (and the initial distance between the two cores in all simulations with $e_i \le 0.7$), $a_{\text{i}}$ is the initial semi-major axis of the binary system, $e_{\text{i}}$ is the initial eccentricity and $R_{1,\text{RL}}$ is the Roche-Lobe radius of the giant, at the orbit's apocenter ($r_{\text{i}}^{\text{a}}$).}
\end{table*}

\subsection{Grazing the envelope at the pericenter}
Our first simulated system contained a mass ratio which, with 0 initial eccentricity, results in a short period stable binary (Sim. 1R06-0 in table \ref{configurations-peri}). The CE evolution of this system has been studied extensively {\citep[][and more]{Pass+12,Iac+17,Reichardt19,glanz2020triple}}; but even-though it was found and indicated that a small eccentricity has been developed during the CE phase, the affect of an initial eccentricity has not been studied yet.
During the CE the eccentricity is not well defined, but we can still calculate the orbit's "quasi eccentricity", in order to examine its evolution. We define the quasi eccentricity as the eccentricity of an orbit with the maximal orbital separation as the apocenter and the minimal orbital separation as the pericenter, such that:
\[
e_{\text{orb}} = \frac{{r_{\text{orb}}}^{\text{max}} - {r_{\text{orb}}}^{\text{min}}}{{r_{\text{orb}}}^{\text{max}} + {r_{\text{orb}}}^{\text{min}}}
\]
where ${r_{\text{orb}}}^{\text{min}}\approx r_p^{\text{orb}}$ and ${r_{\text{orb}}}^{\text{max}} \approx r_a^{\text{orb}}$ are the minimal and maximal separations between the two cores throughout the orbit.

\begin{figure}
    \centering
    \includegraphics[width=\linewidth,clip]{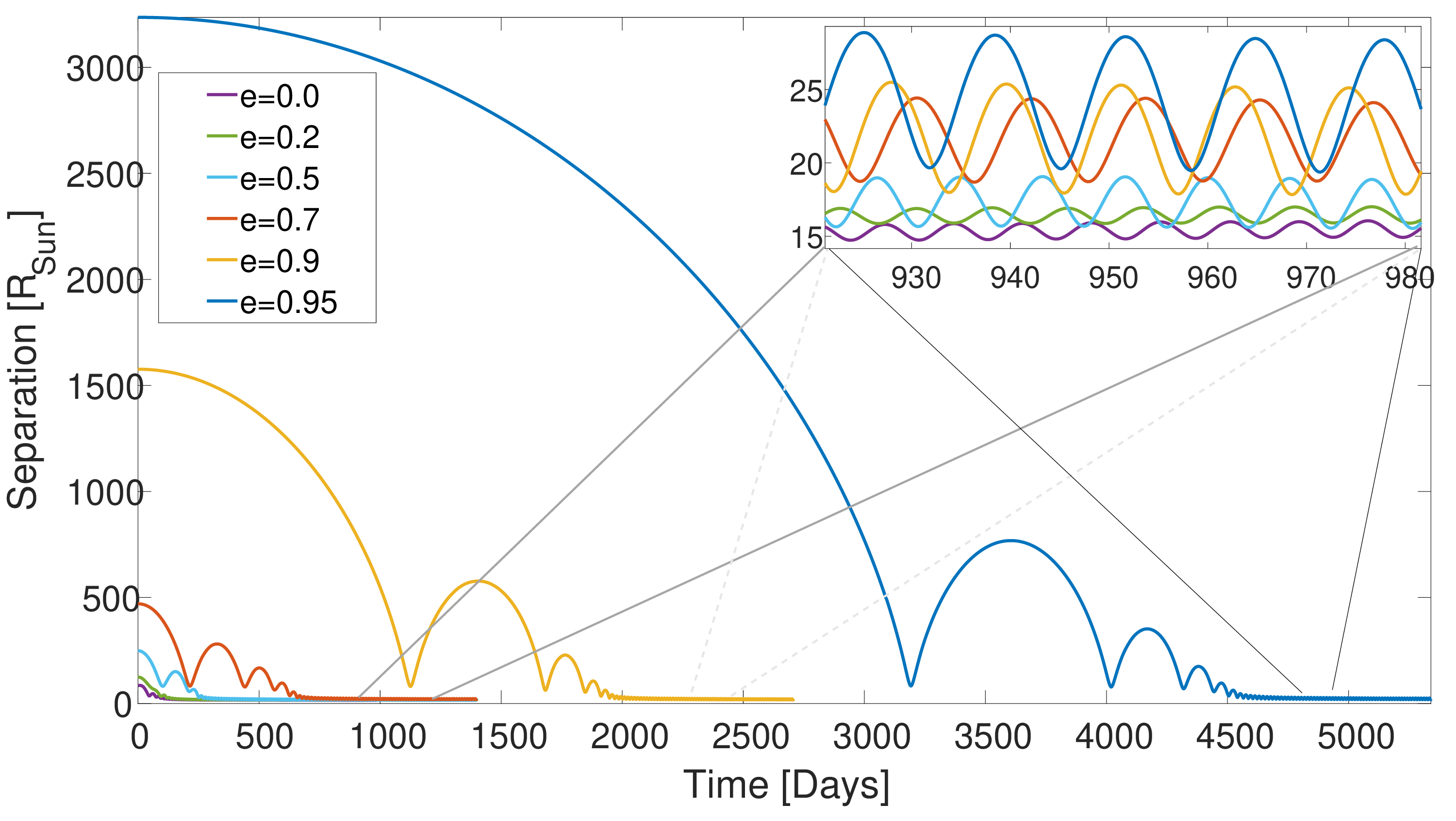}
    \includegraphics[width=\linewidth,clip]{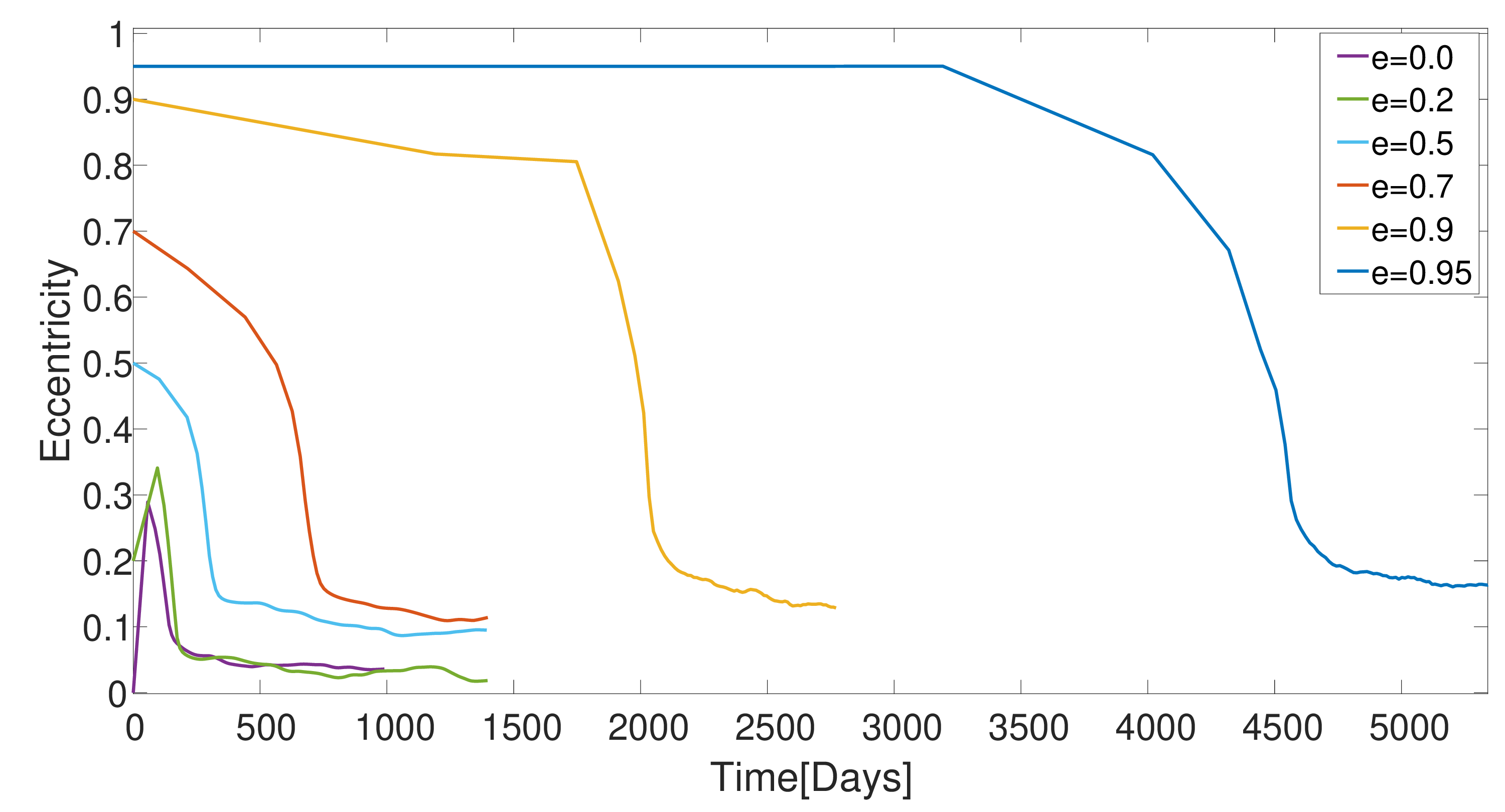}
    \includegraphics[width=\linewidth,clip]{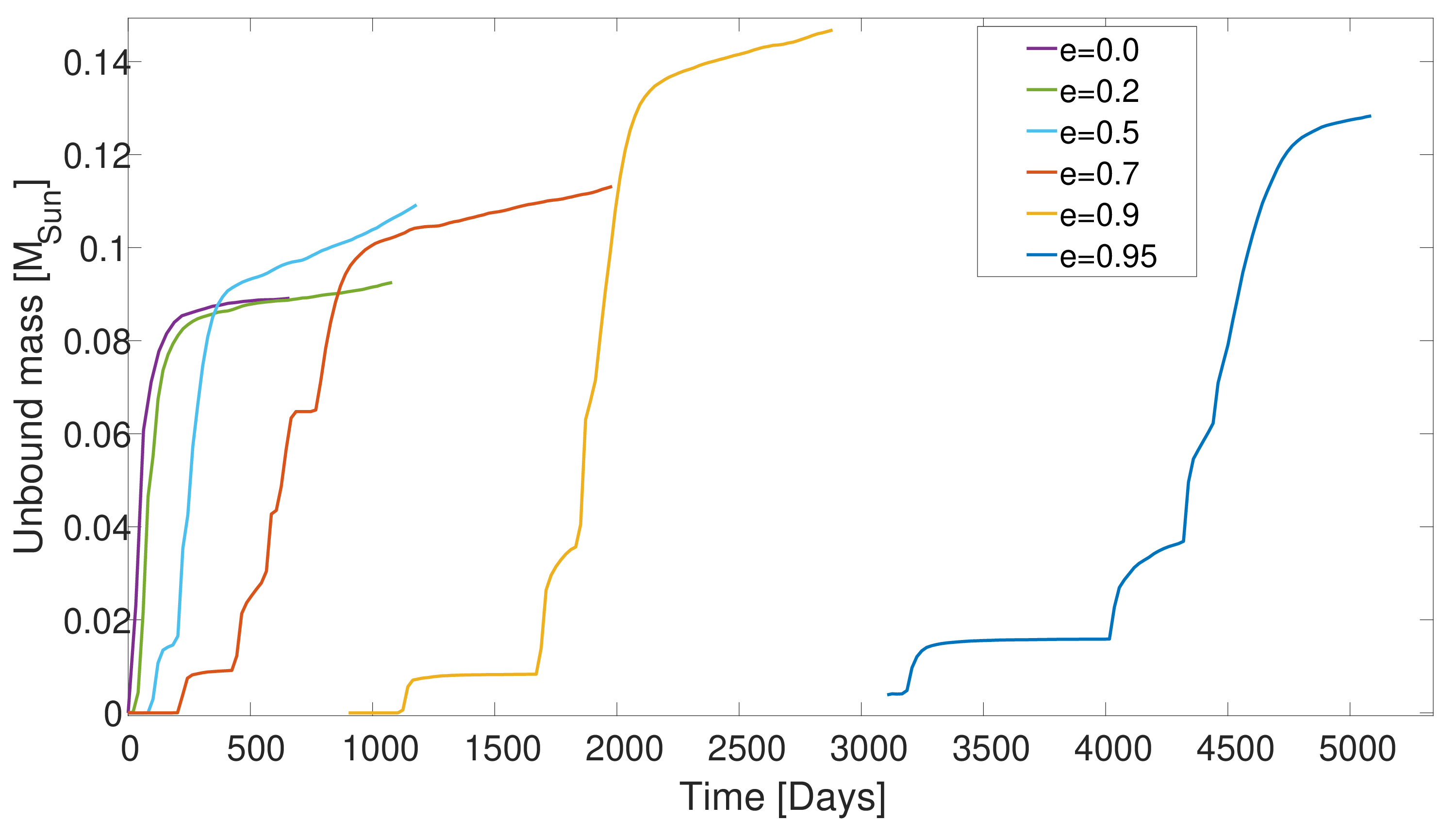}
    \caption{Upper panel: Separation between the primary's core and the companion of simulations with $1M_\odot$ giant and $0.6M_\odot$ companion, initialized with different eccentricities, same periapsis distances and different initial distances (corresponding to their apocenters). Middle panel: eccentricity evolution for same simulations. Bottom: calculated mass loss along the simulations.}
    \label{fig:106p-separation}
\end{figure}

Fig. \ref{fig:106p-separation} presents the evolution of the separations and quasi eccentricities of simulation 106-0,106P2-106P95. All of the orbits circularized significantly, but retain larger final eccentricities for larger initial eccentricities, up to $e_f \approx 0.18$. In addition, as can be seen in the zoomed-in part of the upper panel, systems with larger initial eccentricities show larger final separation.

\begin{figure}
    \centering
    \includegraphics[width=\linewidth,clip]{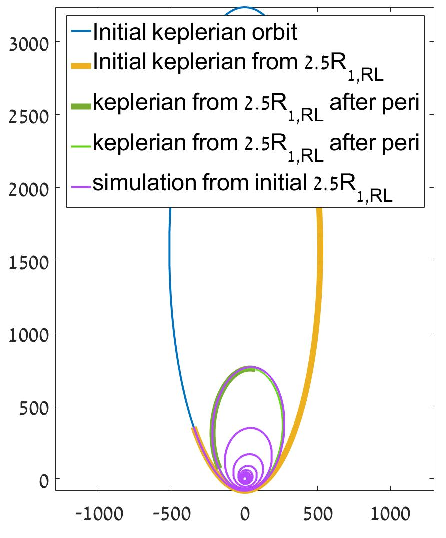}
    \caption{Orbit test for simulation 1R06P95 (Tab. \ref{configurations-peri}) that began at a separation of $2.5R_{1,RL}$. Lines representing the different orbital separations between the companion and the giant's core, for corresponding Keplerian orbits, and the results from the simulation. See Subsec. \ref{initializationInRL} and \ref{stability} for more details.}
    \label{fig:095test}
\end{figure}

While simulation 106-0, 106P2-106P7 began with the orbit at the apocenter, simulations 106P9 and 106P95, that have a large part of low interaction between the companion and the gas, were initialized at separations of $4R_{1,RL}$ and $2.5R_{1,RL}$ of the orbits (see Subsec. \ref{initializationInRL}), with larger softening lengths for the point mass particles. The verification of these extreme eccentric orbits were done as described in Subsec. \ref{lowInteractionTests}, and is presented for 106P96 in Fig. \ref{fig:095test}.

\begin{figure}
    \centering
    \includegraphics[width=\linewidth,clip]{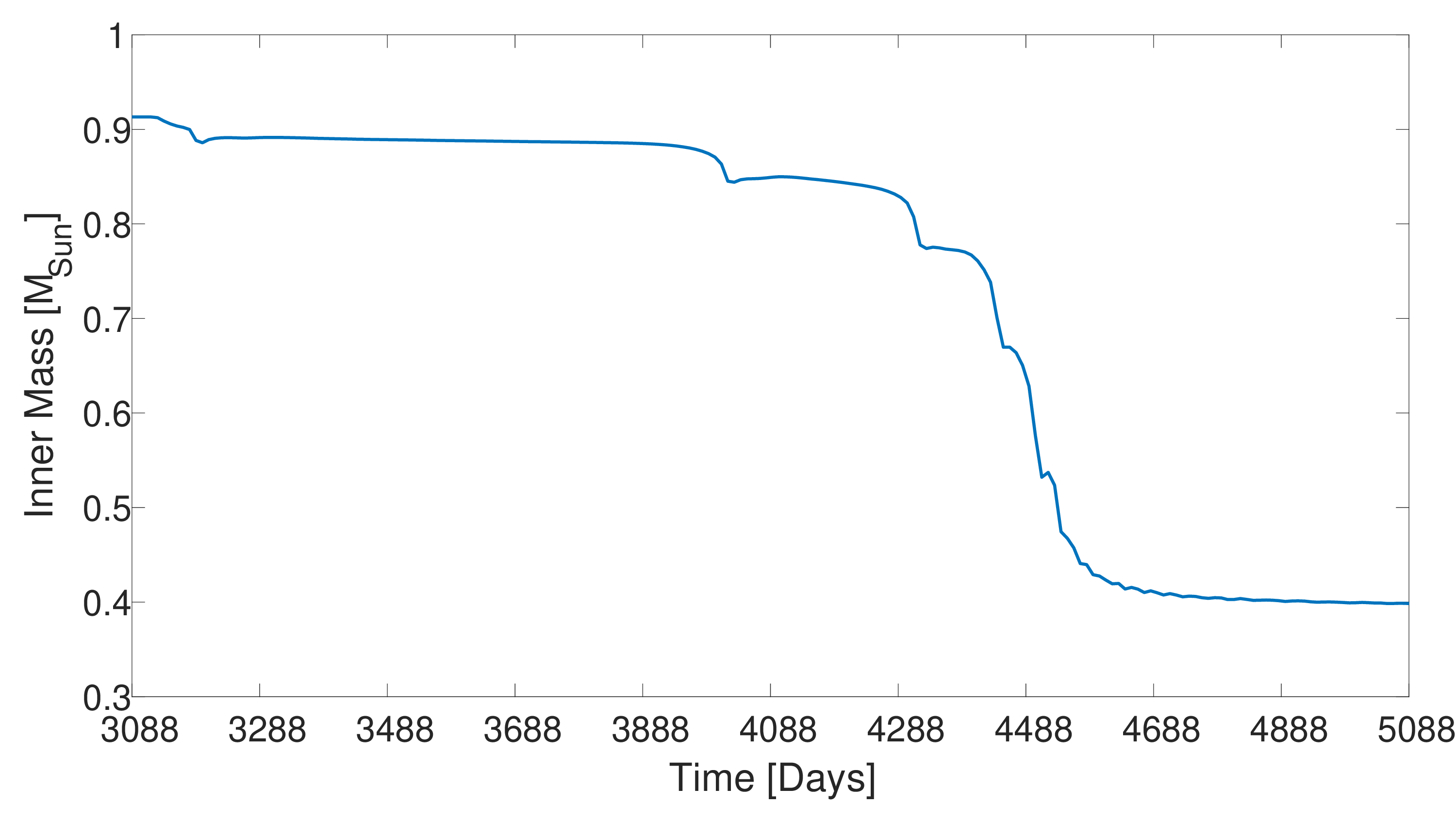}
    \includegraphics[width=\linewidth,clip]{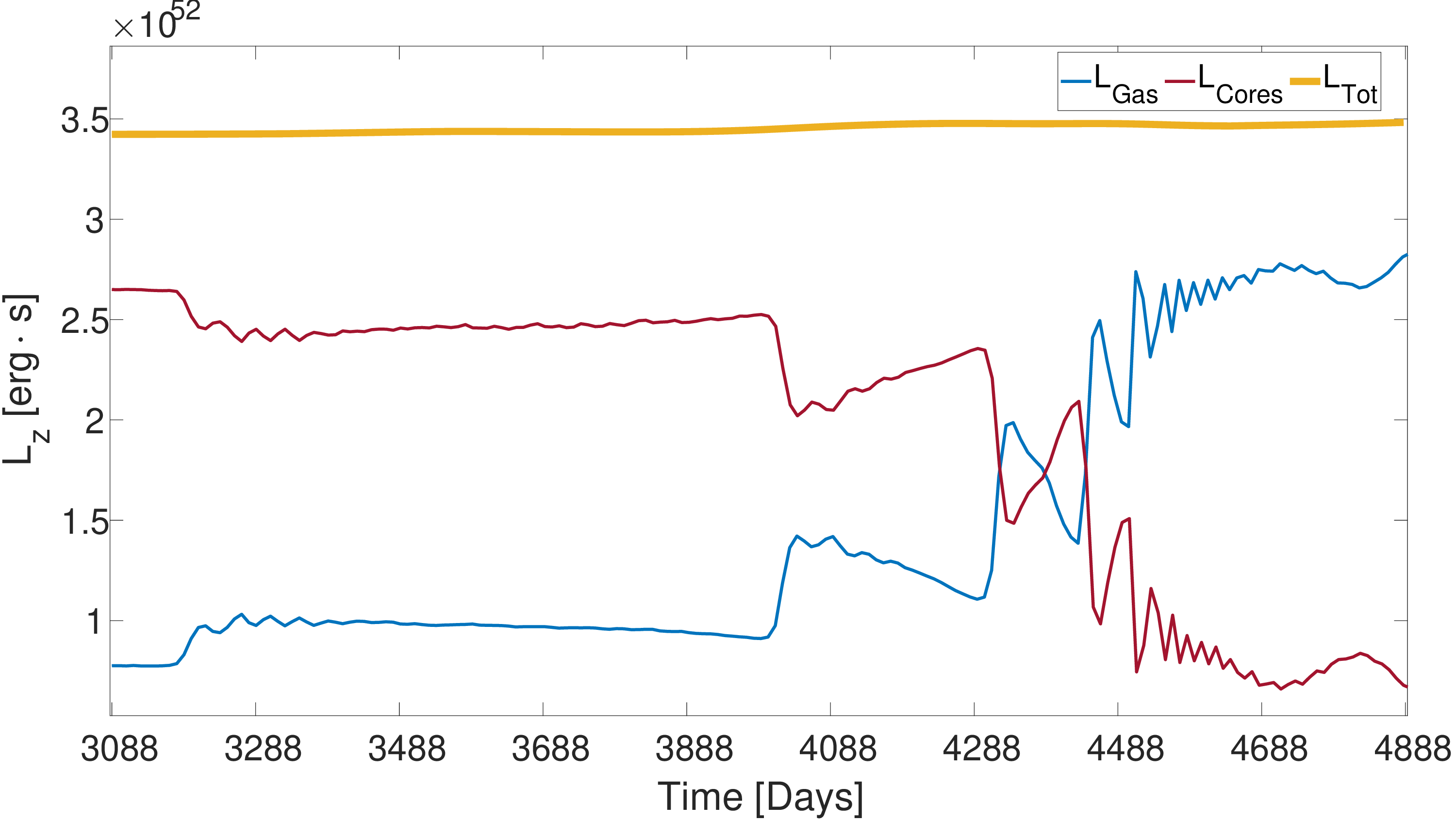}
    \includegraphics[width=\linewidth,clip]{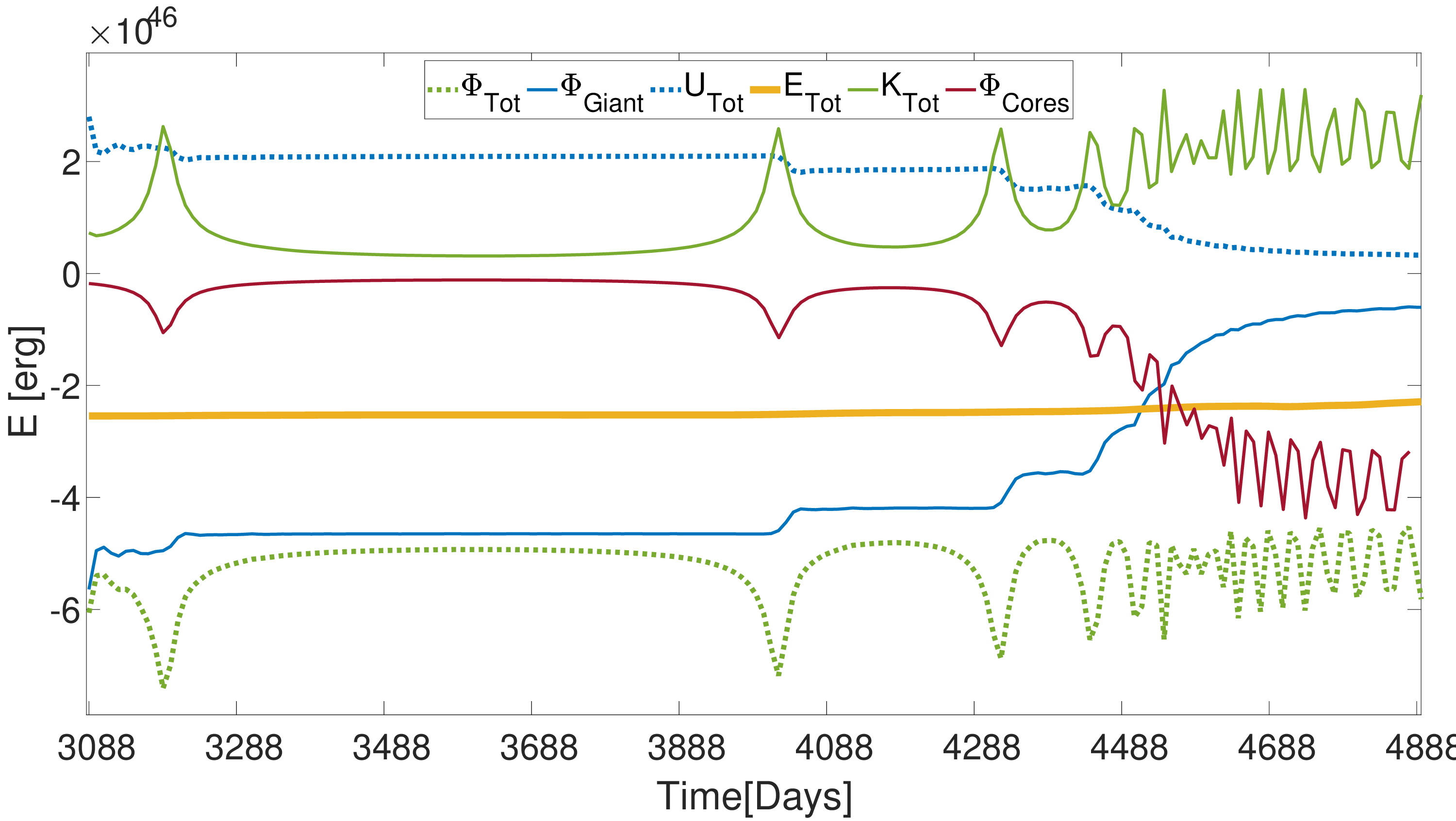}
    \caption{Upper panel: Mass enclosed within the region between the primary's core and the companion (including the mass of the core) of simulation 1R06P95 with $1M_\odot$ giant and $0.6M_\odot$ companion, initialized with 0.95 initial eccentricity. The time is measured in respect to to the initialization of the hydro simulation with separation of $2.5R_{1,RL}$, $\approx 3000 days$ from the previous apocenter at the corresponding initial Keplerian orbit (see Fig \ref{fig:095test}).
    Middle panel: Angular momenta for same simulation. Lower Panel: Energy conservation for same simulation.}
    \label{fig:106p-innermass-angular-energy}
\end{figure}

\begin{figure}
	\begin{tabular}{|c|c|c|}
	\multicolumn{3}{|c|}{0.0 Days} \tabularnewline
	\multicolumn{3}{|c|}{\includegraphics[trim=80 0 22 30,clip,height=0.4\columnwidth]{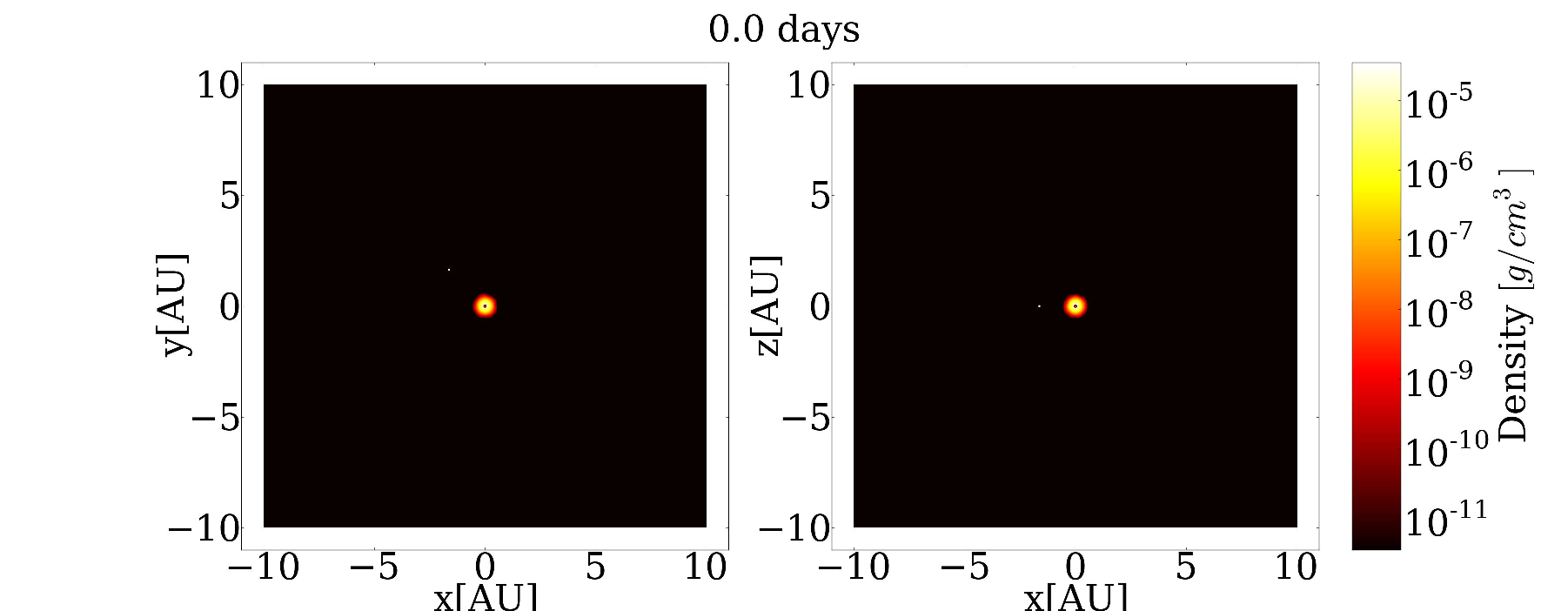}}\tabularnewline
	\multicolumn{3}{|c|}{909.0 Days} \tabularnewline
	\multicolumn{3}{|c|}{\includegraphics[trim=80 0 22 30,clip,height=0.4\columnwidth]{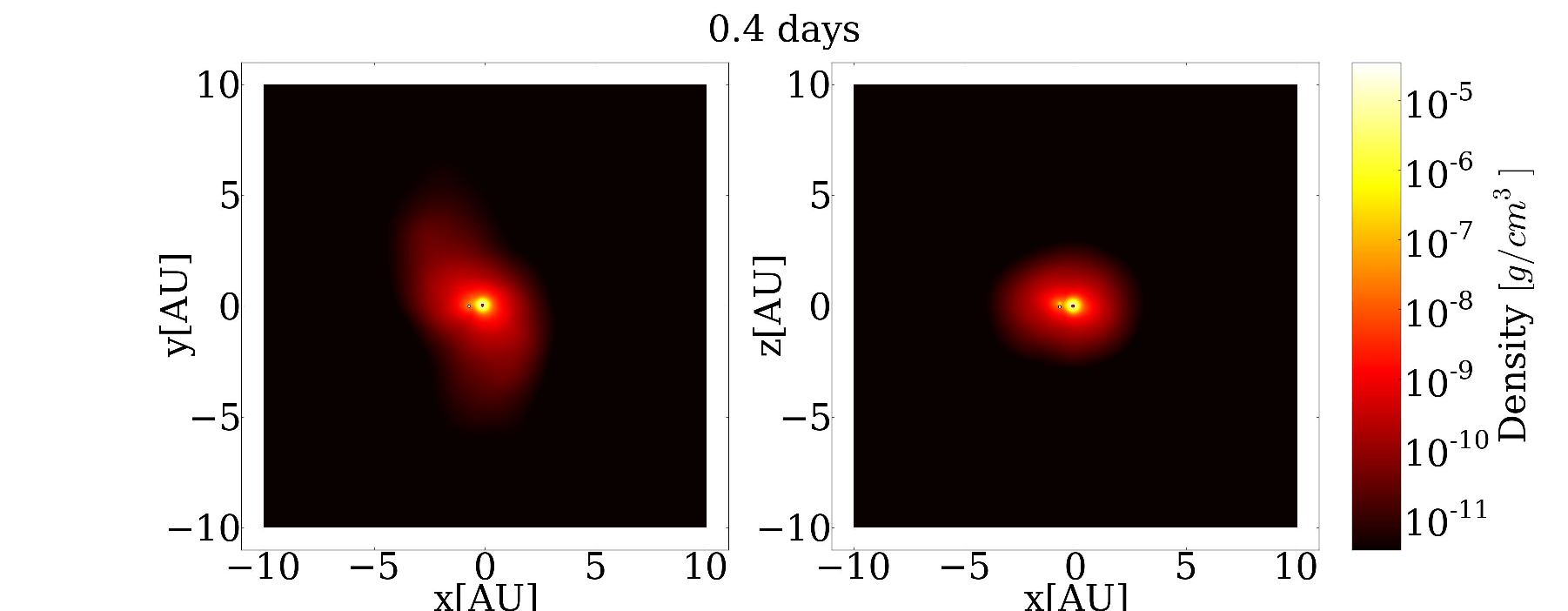}}\tabularnewline
	\multicolumn{3}{|c|}{1244.0 Days} \tabularnewline
	\multicolumn{3}{|c|}{\includegraphics[trim=80 0 22 30,clip,height=0.4\columnwidth]{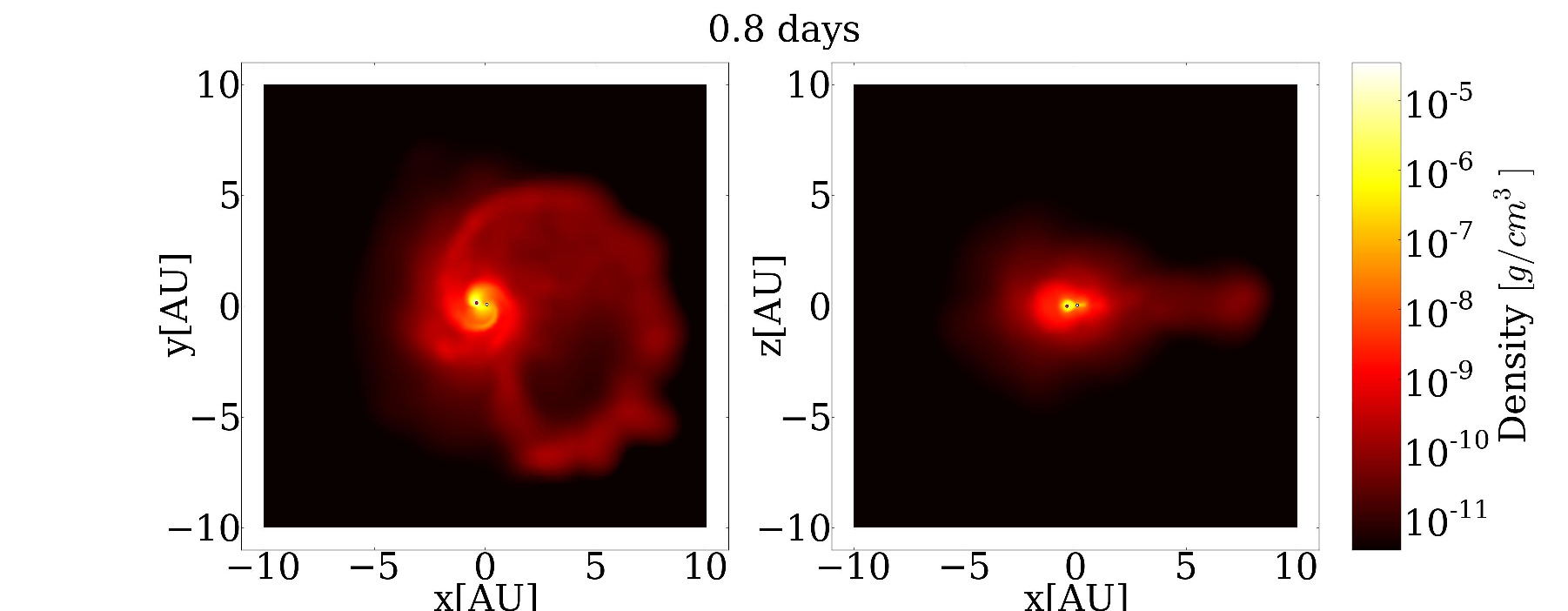}}\tabularnewline
	\multicolumn{3}{|c|}{1425.0 Days} \tabularnewline
	\multicolumn{3}{|c|}{\includegraphics[trim=80 0 22 30,clip,height=0.4\columnwidth]{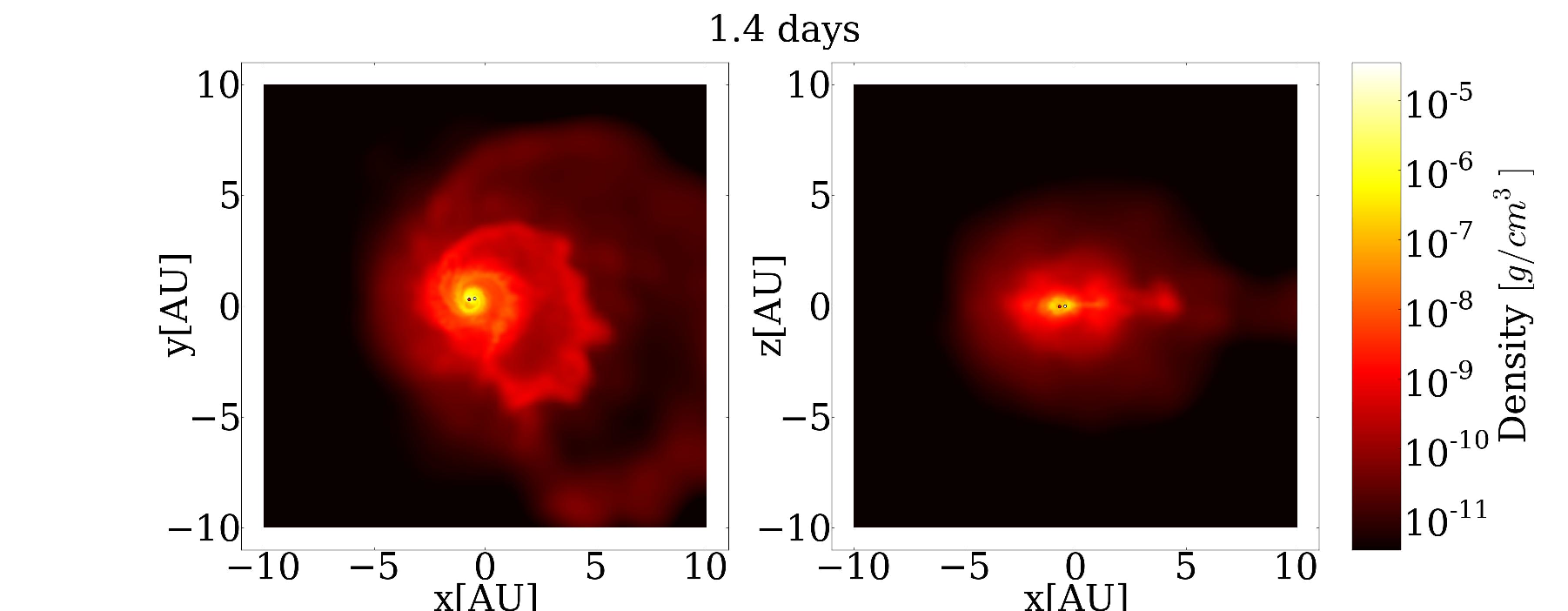}}\tabularnewline
	\multicolumn{3}{|c|}{1655.0 Days} \tabularnewline
	\multicolumn{3}{|c|}{\includegraphics[trim=80 0 22 30,clip,height=0.4\columnwidth]{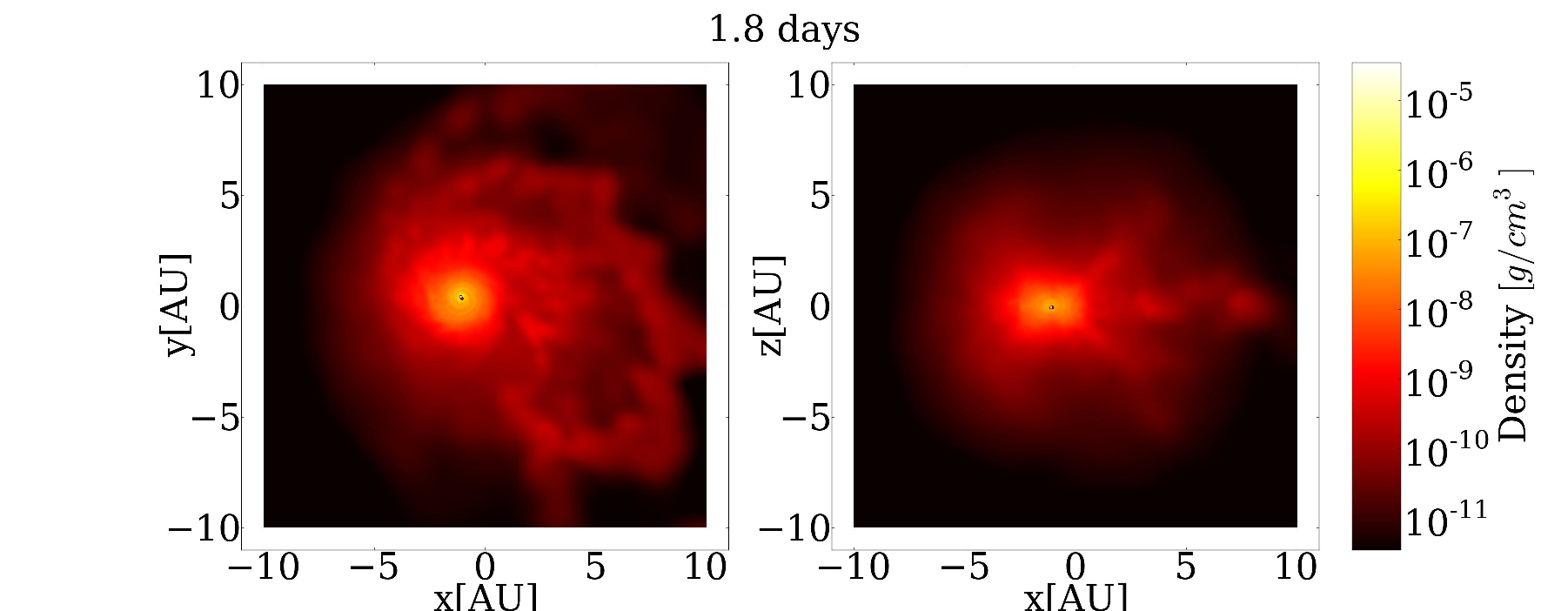}}\tabularnewline
	\end{tabular}
	\caption{\label{fig:snapshots} Simulation 1R06P95 in Tab. \ref{configurations-peri} - a common envelope evolution of a $1M_{\odot}$
		giant with an envelope radius of $83R_\odot$, orbited by a $0.6M_{\odot}$ companion at a separation of $2.5R_{1,RL}$ within an orbital eccentricity of $0.95$. The left and right panels correspond to a plane-parallel view (anti-clock-wise orbit) and an perpendicular view (the companion moves towards the reader), respectively. The companion is marked with white symbol, and the giant
		core is marked by a red  symbol. The symbol sizes do not correspond
		to the stellar sizes and are just shown for clarity. The presented times are from the beginning of the simulation, with a separation of $2.5R_{1,RL}$ between the cores.  }
\end{figure}

Along the spiral-in, the tidal and dynamical friction effect lead to angular momentum transfer from the binary to the gaseous envelope (Fig. \ref{fig:106p-innermass-angular-energy}, upper panel). As a consequence, even when the separation between the two cores increases (after a pericenter passage), the mass located around the core up to the distance of the companion keeps decreasing, while extracting angular momentum and energy from the binary (see Fig. \ref{fig:106p-innermass-angular-energy}). When both the angular momentum of the envelope and its potential energy become larger than the one of the two cores, the system begins to synchronize, and when the inner mass does not change anymore, the dynamical friction goes to zero, the mass unbinding effectively terminates (during the simulation time). The ejection of the material from the inner regions and the effective termination of the orbital dissipation lead to the stabilization of the orbit at a fixed SMA and orbital eccentricity. Density snapshots of simulation 1R06P95 are shown in Fig. \ref{fig:snapshots}, where one can now identify the episodic mass ejections during pericenter passages.

\begin{figure}
    \centering
    \includegraphics[width=\linewidth,clip]{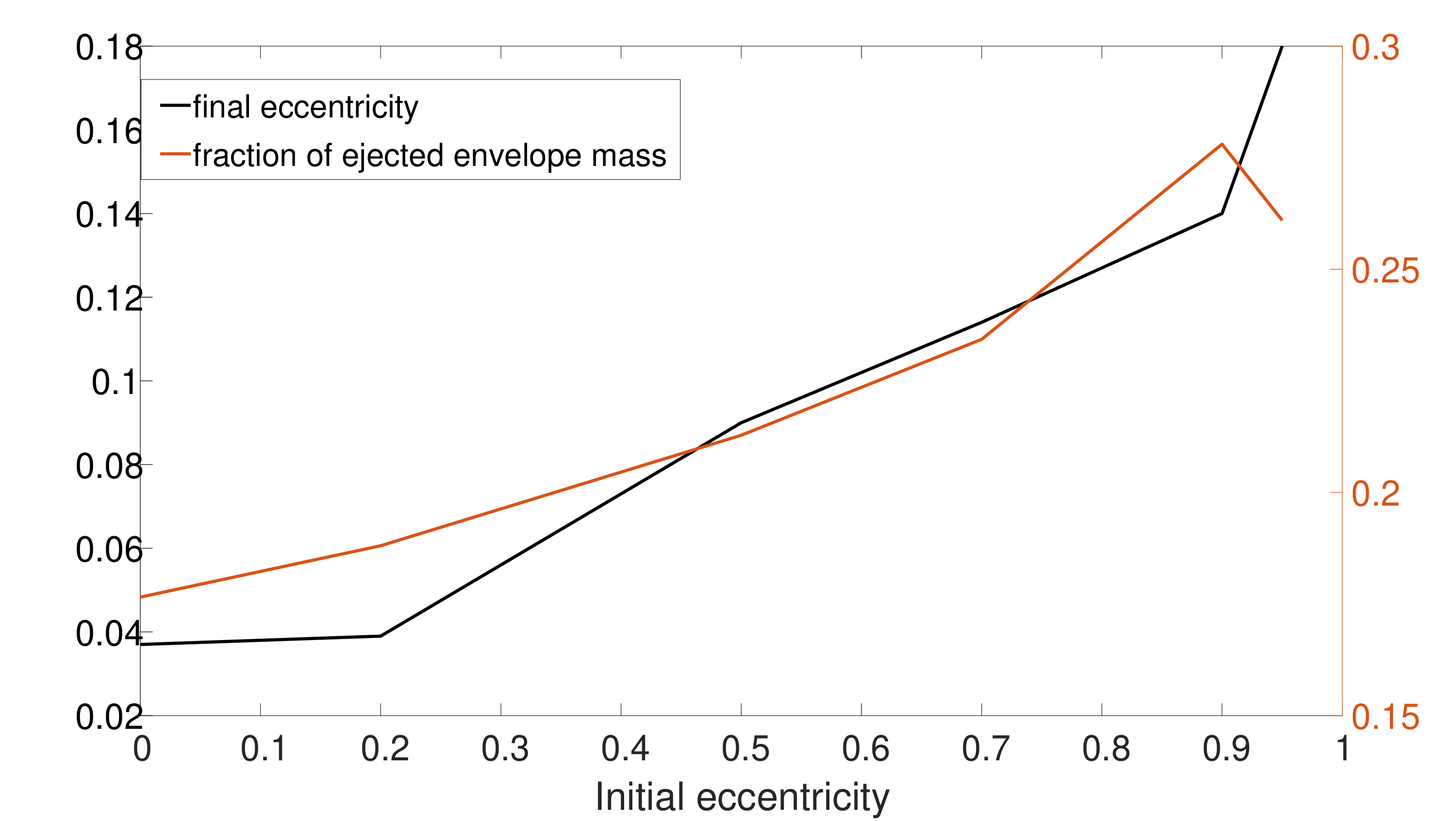}
    \caption{final eccentricities (black, left vertical axis) and the fraction of the unbounded envelope (orange, right vertical axis) versus the initial eccentricity of the system, of models 1R06-0, 1R09P2 to 1R06P95.}
    \label{fig:106p-massloss-eccentricity}
\end{figure}

Due to the higher initial values of the energy and angular momentum, systems with larger eccentricities show larger amount of unbound mass, and as expected from the calculation done by \citet{soker-eccentricitylossinagb}, retain larger eccentricities by the end of the spiral in, despite the strong tidal circularization. Fig. \ref{fig:106p-massloss-eccentricity} shows both final eccentricities and mass loss percentage versus the initial eccentricities, even-though the calculation of mass loss is not accurate (see discussion in Subsec. \ref{massloss}), we can identify a strong correlation between these values and the initial eccentricities. We note that the envelope continues to unbind after the termination of the fast inspiral, but this further mass loss was not calculated due to higher resolution required for longer simulations. In addition, simulation 1R06P95 had a slightly more massive and bigger core, as well as a resulting different smoothing length profile, which can affect the amount of ejected mass.

When calculating the corresponding alpha and gamma CE parameters corresponding to a complete ejection of the envelope in these models \citep{1988ApJliviosoker,2000A&ANelemansGamma,2005MNRASNelemantBetterGamma}, and comparing the final eccentricities according to same constants, we get a good agreement between simulations which began with a giant not yet filling its Roche-Lobe at semi-major axis (1R06P7-1R06P95). However, systems with lower initial eccentricities, which ended with lower final eccentricities, should have increased their eccentricities by this relation.

\begin{figure}
    \centering
    \begin{tabular}{|c|c|c|}
    \hline
    \multicolumn{3}{|c|}{1R06-0}\tabularnewline
    \multicolumn{3}{|c|}{\includegraphics[trim=80 0 22 30,clip,height=0.4\columnwidth]{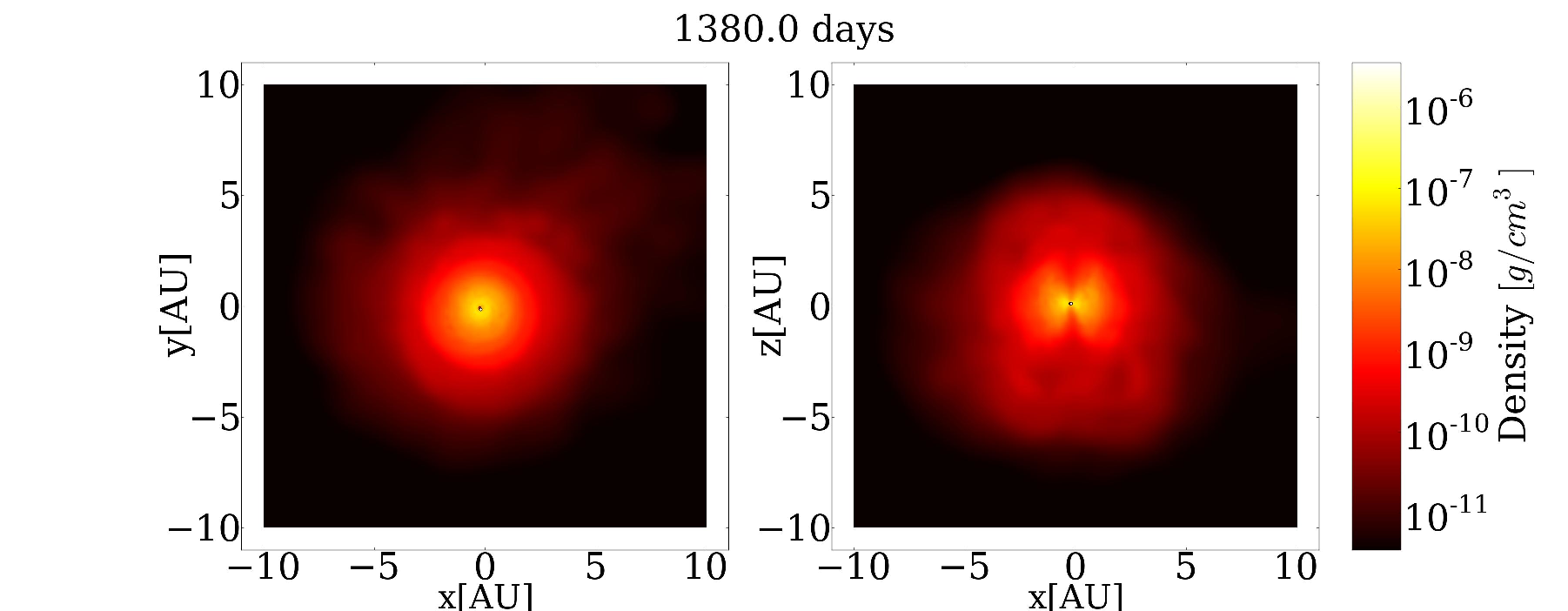}}\tabularnewline
    \hline
    \multicolumn{3}{|c|}{1R06P7}\tabularnewline
    \multicolumn{3}{|c|}{\includegraphics[trim=80 0 22 30,clip,height=0.4\columnwidth]{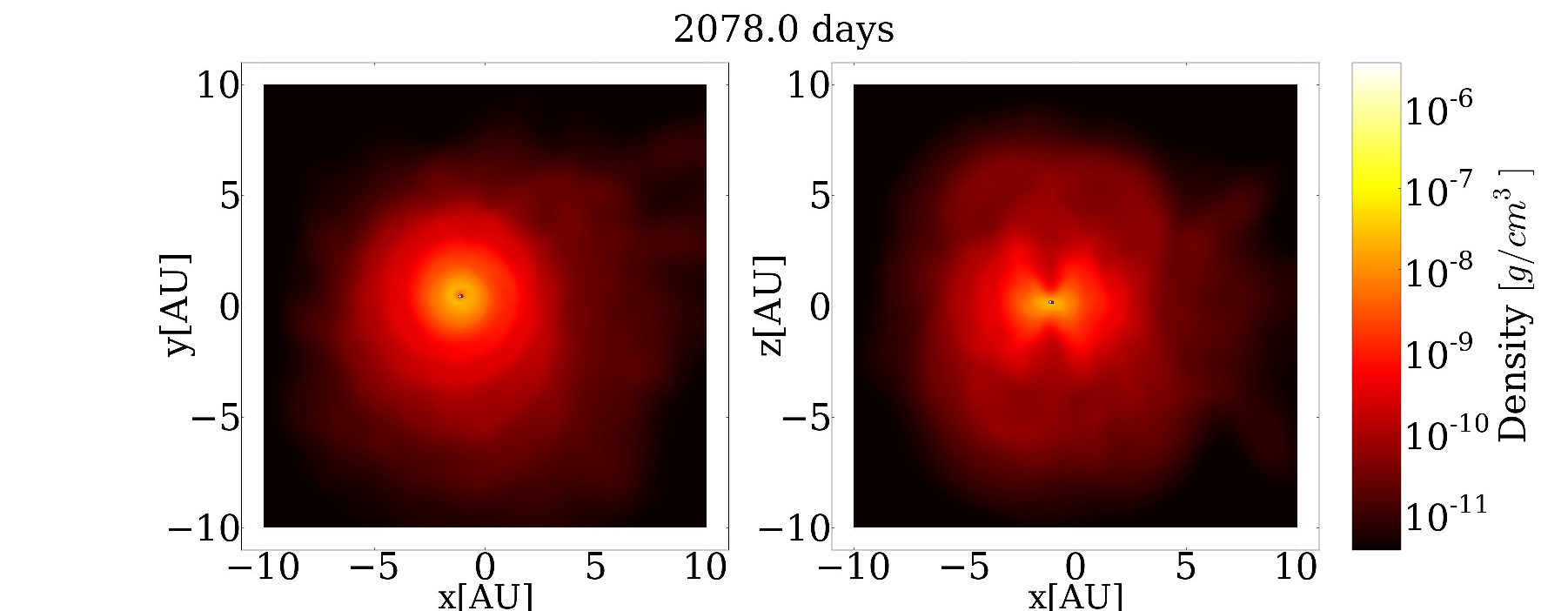}}\tabularnewline
    \hline
    \multicolumn{3}{|c|}{1R06P95}\tabularnewline
    \multicolumn{3}{|c|}{\includegraphics[trim=80 0 22 30,clip,height=0.4\columnwidth]{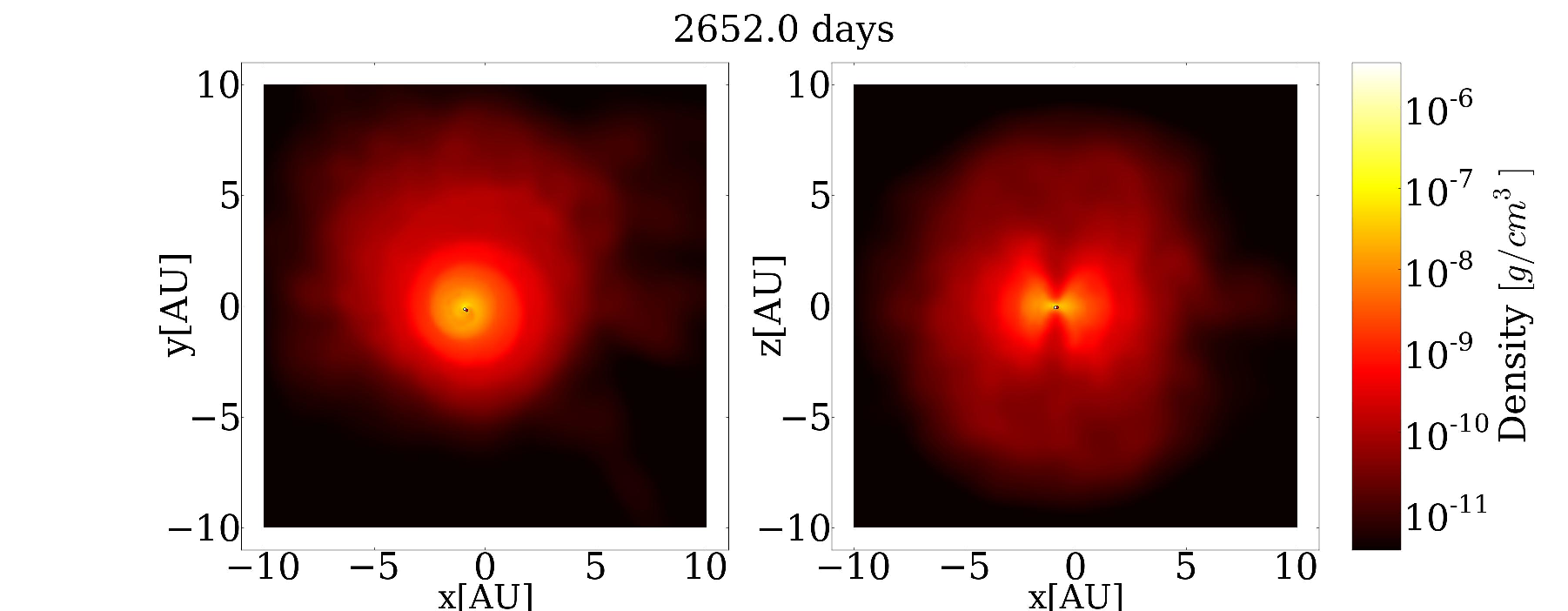}}\tabularnewline
    \hline
    \end{tabular}
    \caption{Final snapshots of simulations 1R206P0, 1R06P7 and 1R06P95.eps}
    \label{fig:1R06P-snapshots}
\end{figure}

In Fig. \ref{fig:1R06P-snapshots}, we show the final snapshots of some of these simulations, suggesting that higher eccentricities lead to an extended shape of the nebula in the direction of the angular momenta. However, this is not the final shape since the material continues to expand after our simulations terminated. 

\begin{figure}
    \centering
    \includegraphics[width=\linewidth,clip]{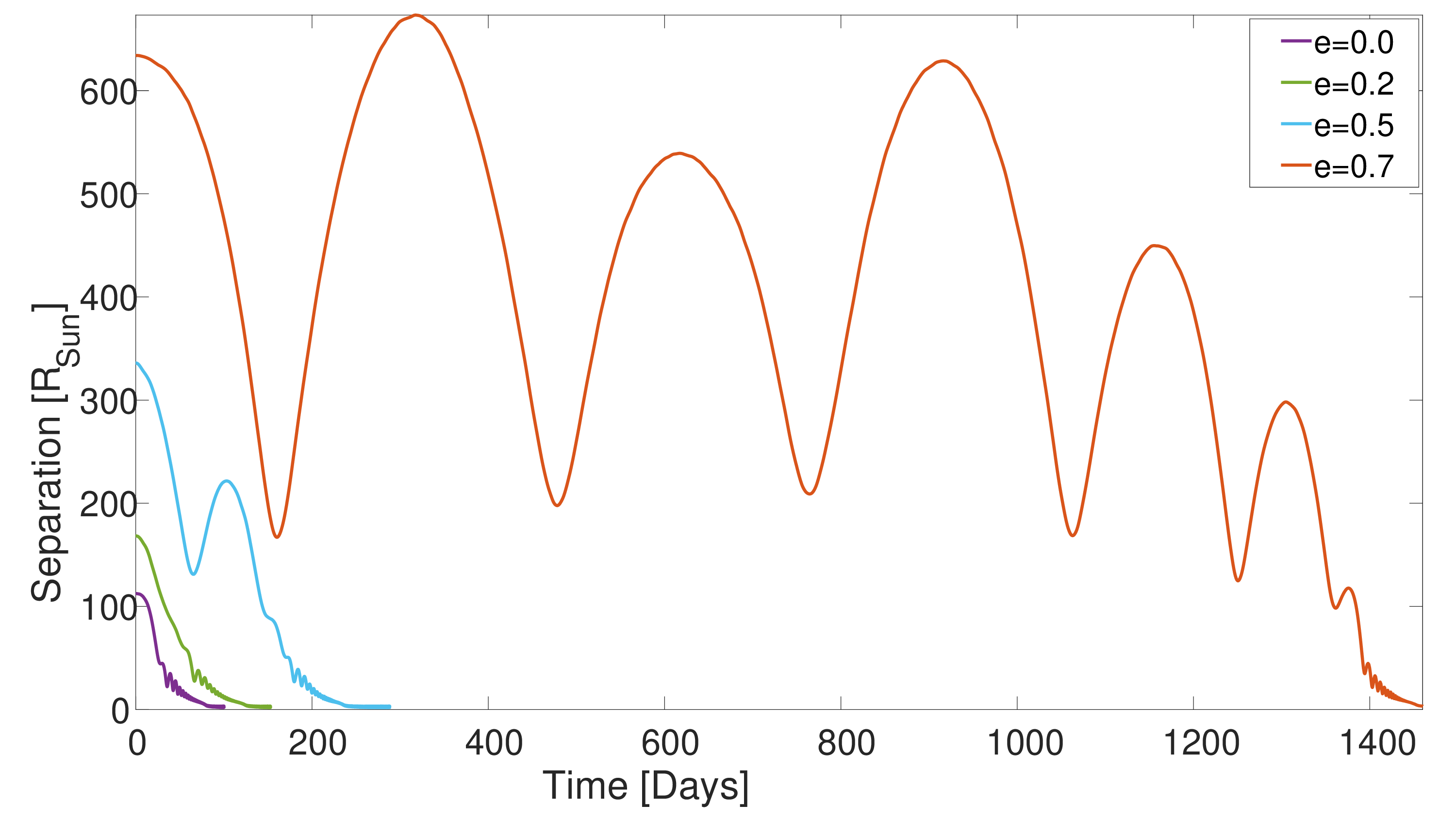}
    \includegraphics[width=\linewidth,clip]{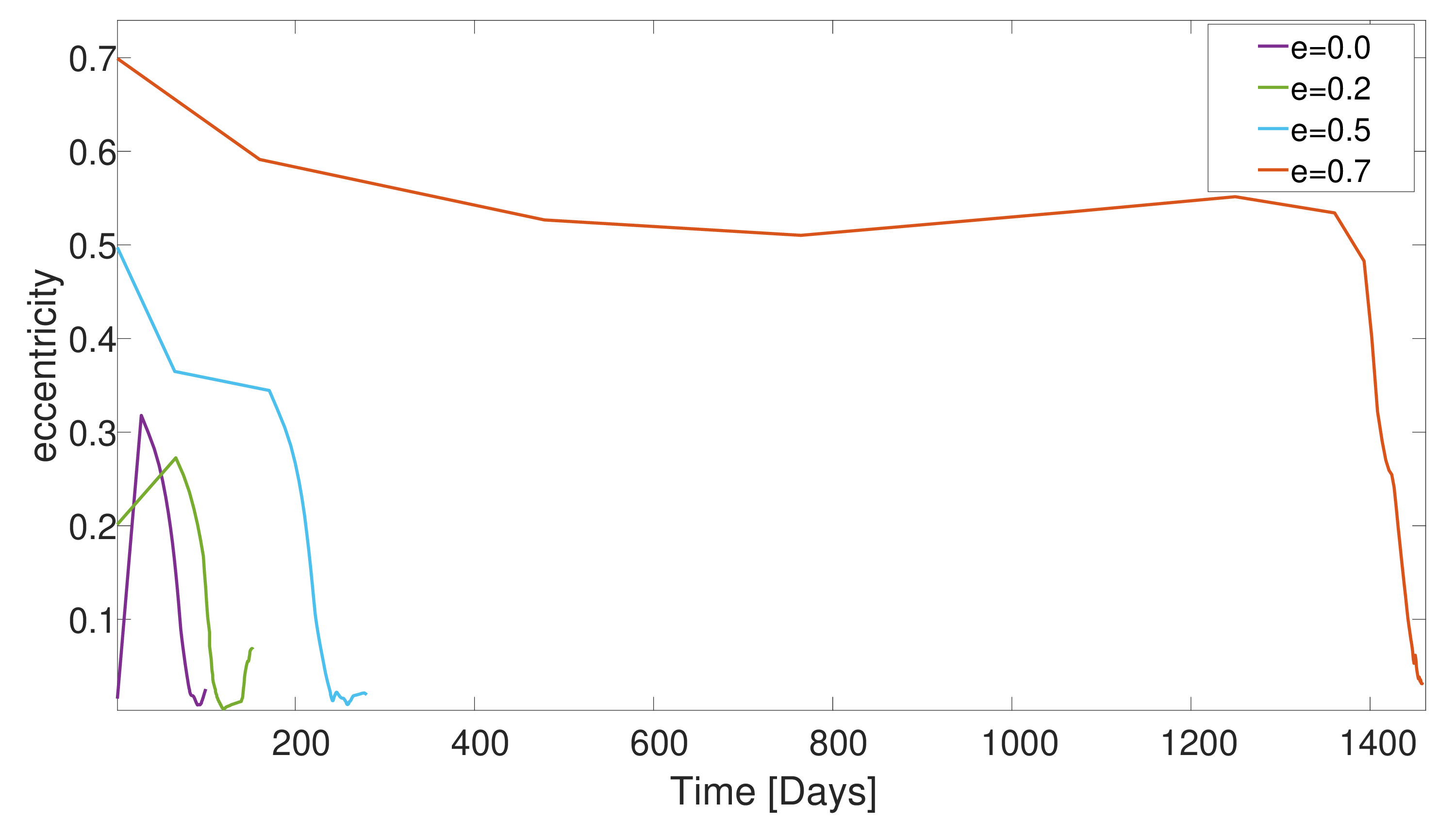}
    \includegraphics[width=\linewidth,clip]{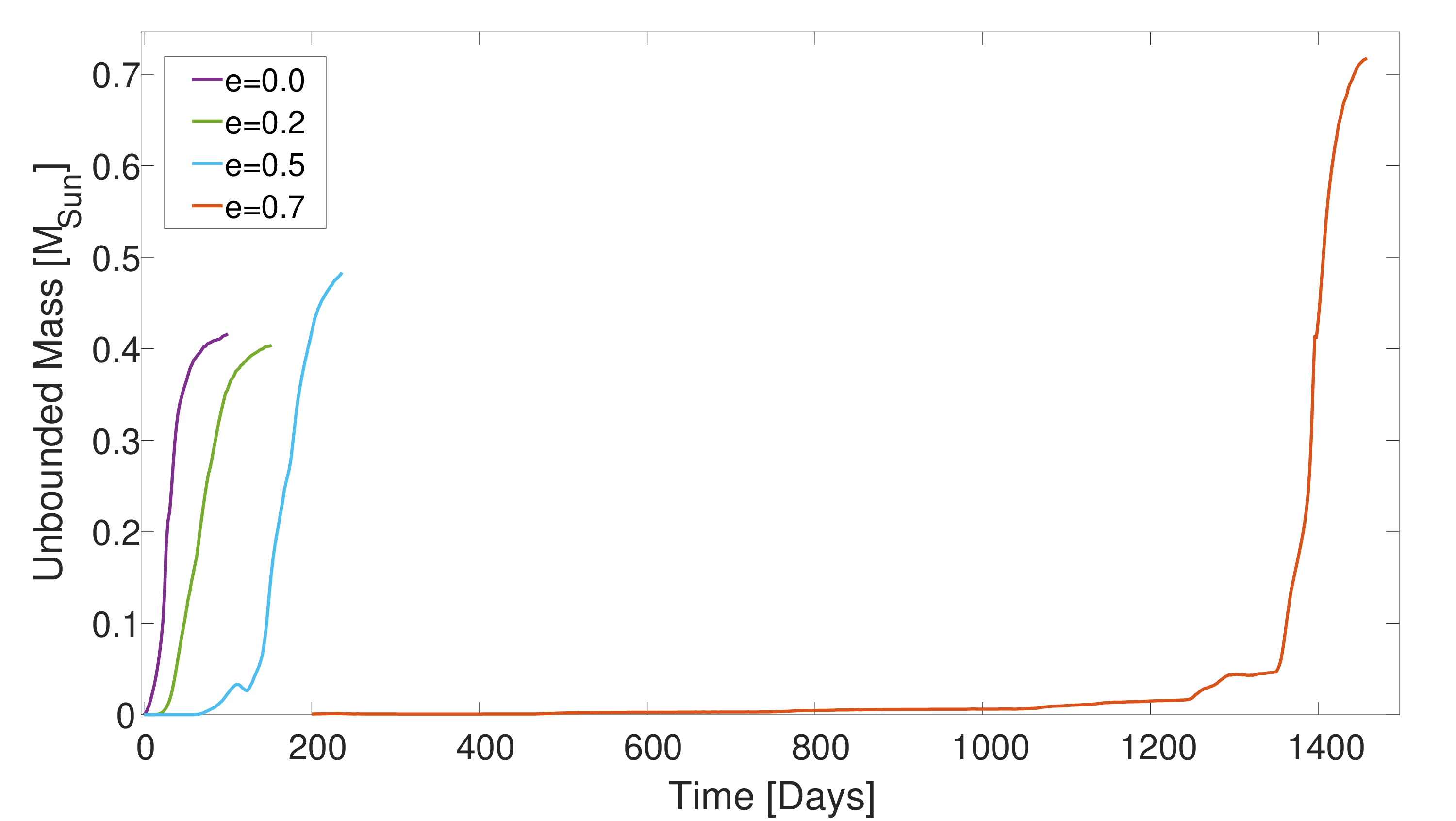}
    \caption{Upper panel: Separation between the primary's core and the companion in simulations of systems with a $8M_\odot$ giant and a $2M_\odot$ companion, initialized with different eccentricities, same periapsis distances and different initial separations (corresponding to their apocenters). Middle panel: eccentricity evolution for the same simulations. Bottom: Unbound mass during the simulations.}
    \label{fig:82g-separation}
\end{figure}

 The second system we test has an $8M_\odot$ Red giant as a primary, and $2M_\odot$ companion (simulations 8R2G0-8R2G7). All of these simulations ended with a core merger (as defined in \citealt{glanz2020triple}), but as shown in Fig. \ref{fig:82g-separation}, still highly circularized prior to this merger. The mass loss corresponding to these points can be seen in Tab. \ref{results-table}, showing larger amount of ejected, unbound mass found for systems with larger initial eccentricities.

The shape of the final ejecta in these cases cannot be studied without modeling the actual merger. Any asymmetry in the mass ejection that is present during the inspiral may disappear later on, due to the energy that is released from the merger \citep{2016PejchaMergerEjecta}. 

 We simulated another giant with a middle range mass of about $4M_\odot$, which as shown in Fig. \ref{fig:4evolution}, has a maximal radius of only $42R_\odot$ during in Red giant stage. In simulations 4R06P0 - 4R06P95 in Tab. \ref{configurations-peri} we explore the CEE of this giant with a $0.6M_\odot$ companion with initial eccentricities ranging up to $e_i=0.95$. Since the mass ratio between the primary and secondary stars is very large, all simulations, including the most extreme cases ended with a core merger. As was done to the previous systems with extreme initial eccentricities, simulations 4R06P9 and 4R06P95 were initialized at locations corresponding to separations of $3R_{1,RL}$.

\begin{figure}
    \centering
    \includegraphics[width=\linewidth,clip]{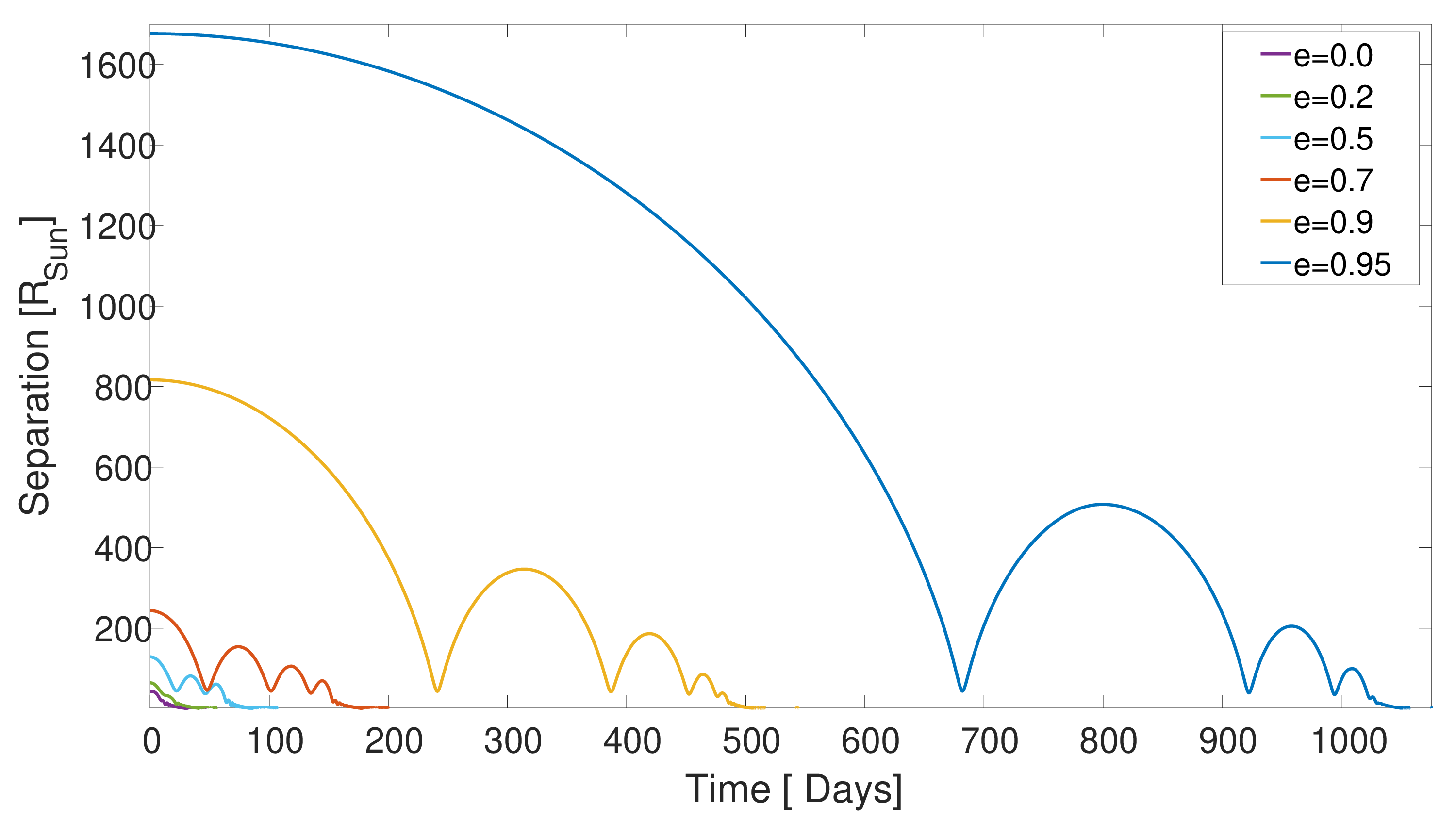}
    \includegraphics[width=\linewidth,clip]{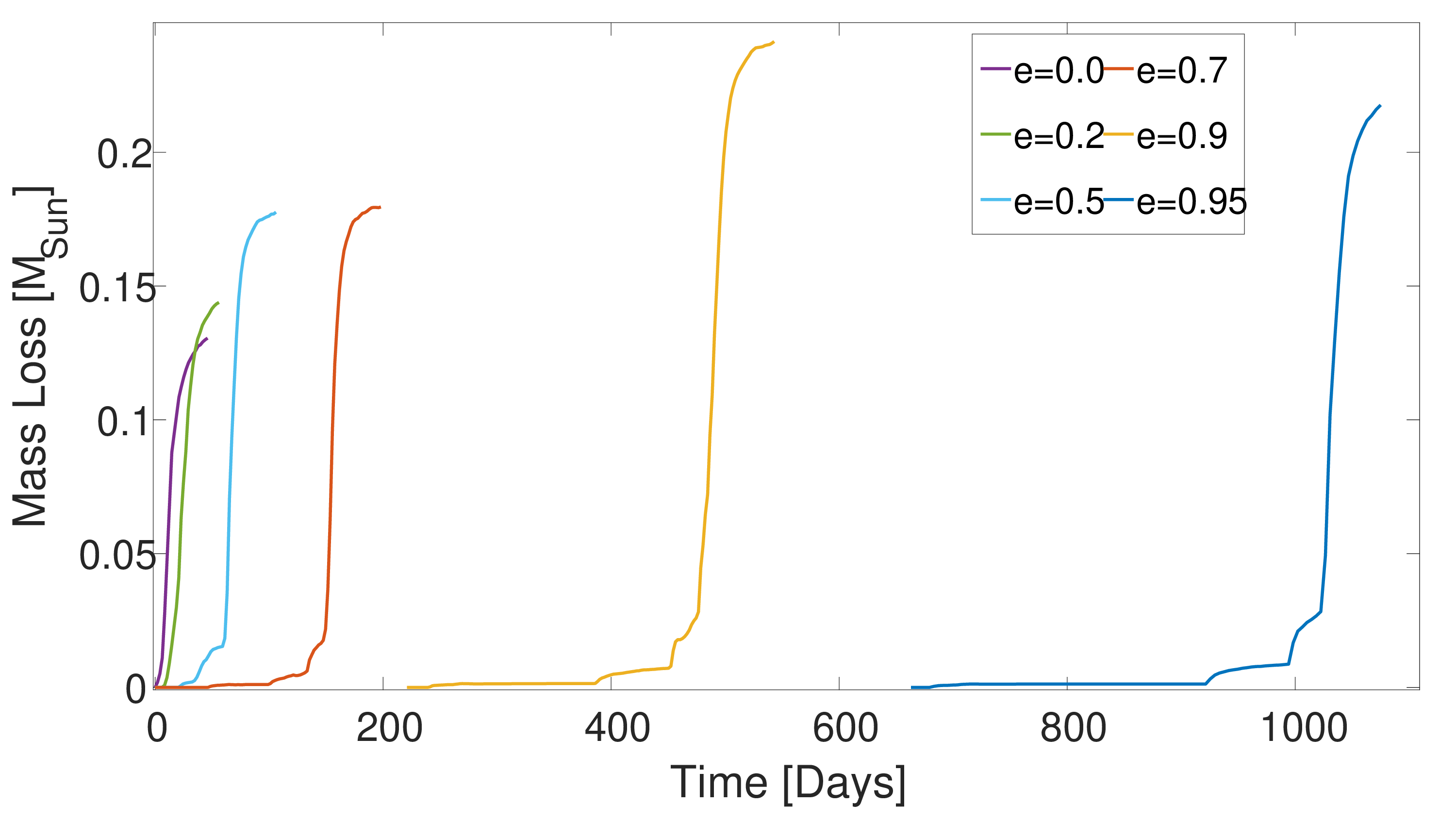}
    \caption{Upper panel: The separation between the primary's core and the companion in simulations of systems with a $4M_\odot$ red giant and a $0.6M_\odot$ companion, initialized with different eccentricities, same pericenter distances and different initial separations (corresponding to their apocenters). Lower panel: mass loss calculated for same simulations.}
    \label{fig:4r06p-separation}
\end{figure}

Due to the large binding energy of this giant, the amount of mass ejected prior to the merger was very low (see Fig. \ref{fig:4r06p-separation}), and the ejecta were therefore almost axial symmetric by that time.

\subsection{Systems with varying pericenters located inside the envelope}
Typically, evolution to the CE phase occurs following the slow evolutionary growth of the primary star, which expands beyond it's Roche-lobe, and strong interactions occur though initially grazing encounters. However, in some systems dynamical evolution could give rise to evolution into close encounters and collisions (i.e. encounters at impact parameters closer the stellar radius). Such systems can form following velocity kicks imparted to neutron-stars/black-holes, through violent dynamical interactions in destabilized triple systems \cite{per+12c,Mic+14,glanz2020triple} or through secular Lidov-kozai \citep{1962Lidov, 1962Kozai} or quasi-secular \citep{Ant+12} evolution in triple stellar systems \citep[e.g.][and references therein]{2009peretsBlueStragglers,per+12c,Sha+13,Mic+14, toonen2016evolution,Nao16, Moe+17}.

In the following, we discuss systems in which the initial peri-center is inside the envelope, unlike the initially grazing systems discussed before. We consider systems with the same apocenter, but different peri-center inside the envelope.

\begin{table*}
\begin{tabular}{|c|c|c|c|c|c|c|c|c|c|}
\hline 
 Sim & $M_\text{1}$ &$M_\text{2}$ &$M_{1,\text{core}}$ &$R_\text{1}$ &$r_{\text{i}}^{\text{a}}$ &$a_{\text{i}}$& $e_{\text{i}}$ & $R_{1,\text{RL}}$ 
\\ &(${\rm M_{\odot}}$)&(${\rm M_{\odot}}$)&(${\rm M_{\odot}}$)&(${\rm R_{\odot})}$&$({\rm R_{\odot})}$ &$({\rm R_{\odot})}$&&$({\rm R_{\odot})}$\tabularnewline
\hline 
\hline 
1R06-0 &1 & 0.6 & 0.388 & $83$ & 83 & 83 & 0.0 & 34  \tabularnewline
1R06A2 &1 & 0.6 & 0.388 & $83$ & 83 & 69 & 0.2 & 34  \tabularnewline
1R06A5 &1 & 0.6 & 0.388 & $83$ & 83 & 55.3 & 0.5 & 34  \tabularnewline
1R06A7 &1 & 0.6 & 0.388 & $83$ & 83 & 49 & 0.7 & 34  \tabularnewline

\hline 
8R2-0 & 8 & 2 & 1.03 & 110 & 216 & $216$ & 0.0 & 108 \tabularnewline
8R2A2 & 8 & 2 & 1.03 & 110 & 216 & 180 & 0.2 & 108 \tabularnewline
8R2A5 & 8 & 2 & 1.03 & 110 & 216 & 144 & 0.5 & 108 \tabularnewline
8R2A7 & 8 & 2 & 1.03 & 110 & 216 & 127 & 0.7 & 108 \tabularnewline
\hline 
\end{tabular}\caption{\label{configurations-apo} Initial configuration of the simulated systems with varying pericenter distances. $M_\text{1}$ is the mass of the primary at zero age in the main sequence, $M_\text{2}$ is the mass of the secondary, $M_{1,\text{core}}$ and $R_\text{1}$ are the core mass and radius of the primary at the beginning of the common envelope, $r_{\text{i}}^{\text{a}}$ is the initial distance between the giant core and the companion and is the apocenter of their mutual orbit, $a_{\text{i}}$ is the initial semi-major axis of the binary system, $e_{\text{i}}$ is the initial eccentricity and $R_{1,\text{RL}}$ is the Roche-Lobe radius of the giant. }
\end{table*}

\begin{figure}
    \centering
    \includegraphics[width=\linewidth,clip]{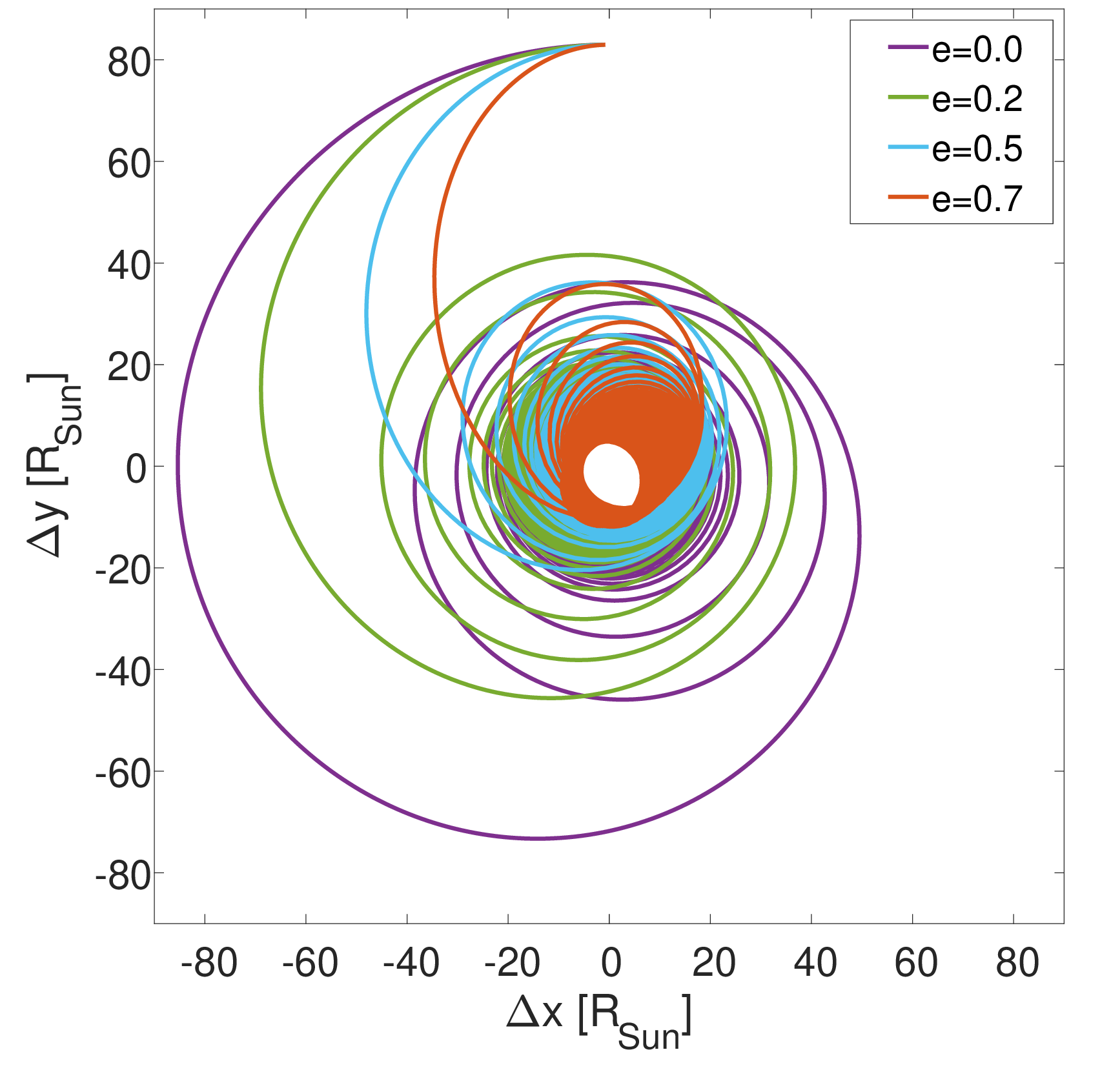}
    \caption{The orbit (distances between the two cores) of simulations with $1M_\odot$ giant and $0.6M_\odot$ companion, initialized with same distance (same apoapsis distance) and different eccentricities. }
    \label{fig:106a-orbit}
\end{figure}

\begin{figure}
    \centering
    \includegraphics[width=\linewidth,clip]{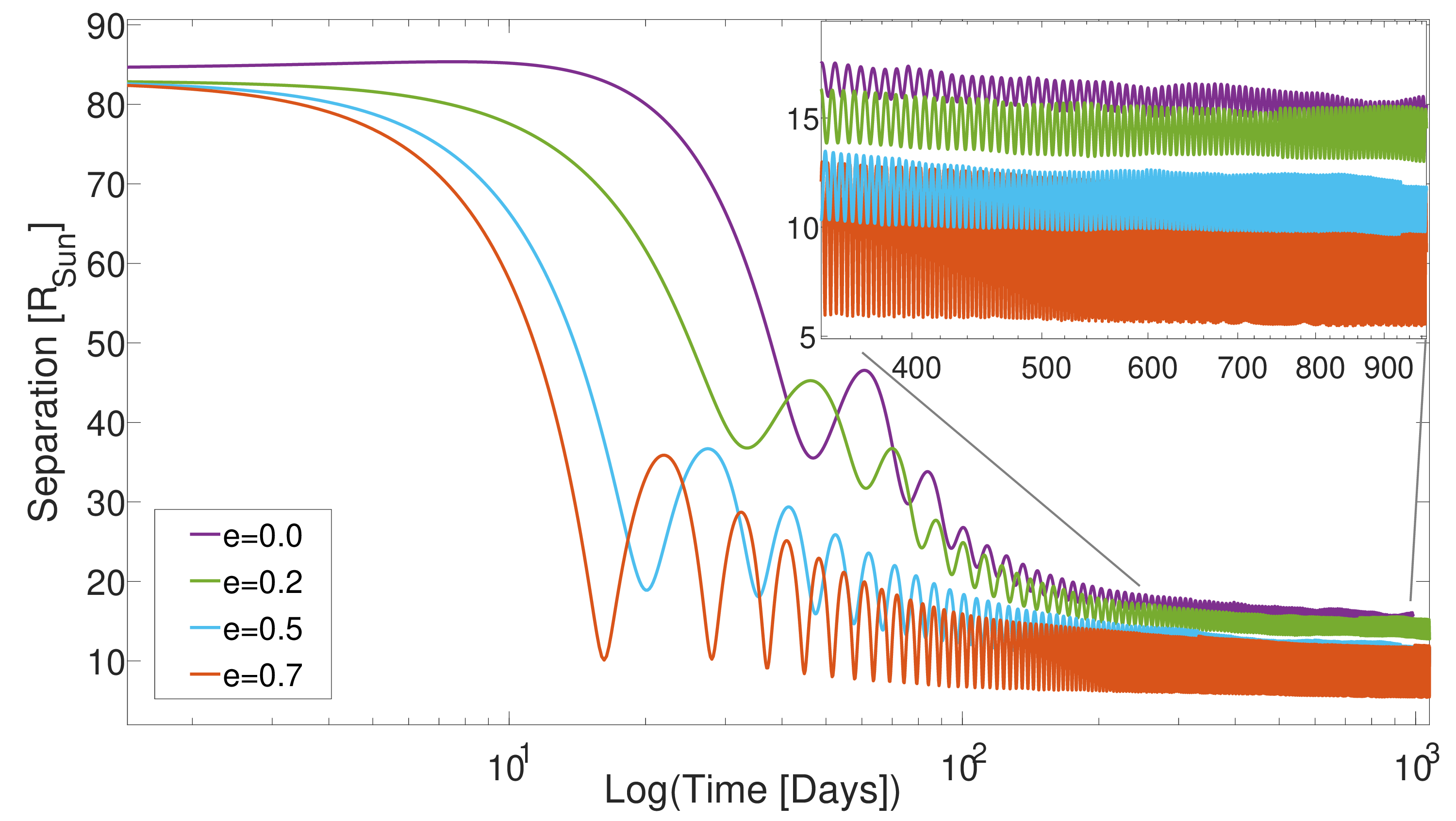}
    \caption{Separation between the primary's core and the companion of simulations with $1M_\odot$ giant and $0.6M_\odot$ companion, initialized with same distance (same apocenter distance) and different eccentricities.}
    \label{fig:106a-separation}
\end{figure}

\begin{figure}
    \centering
    \includegraphics[width=\linewidth,clip]{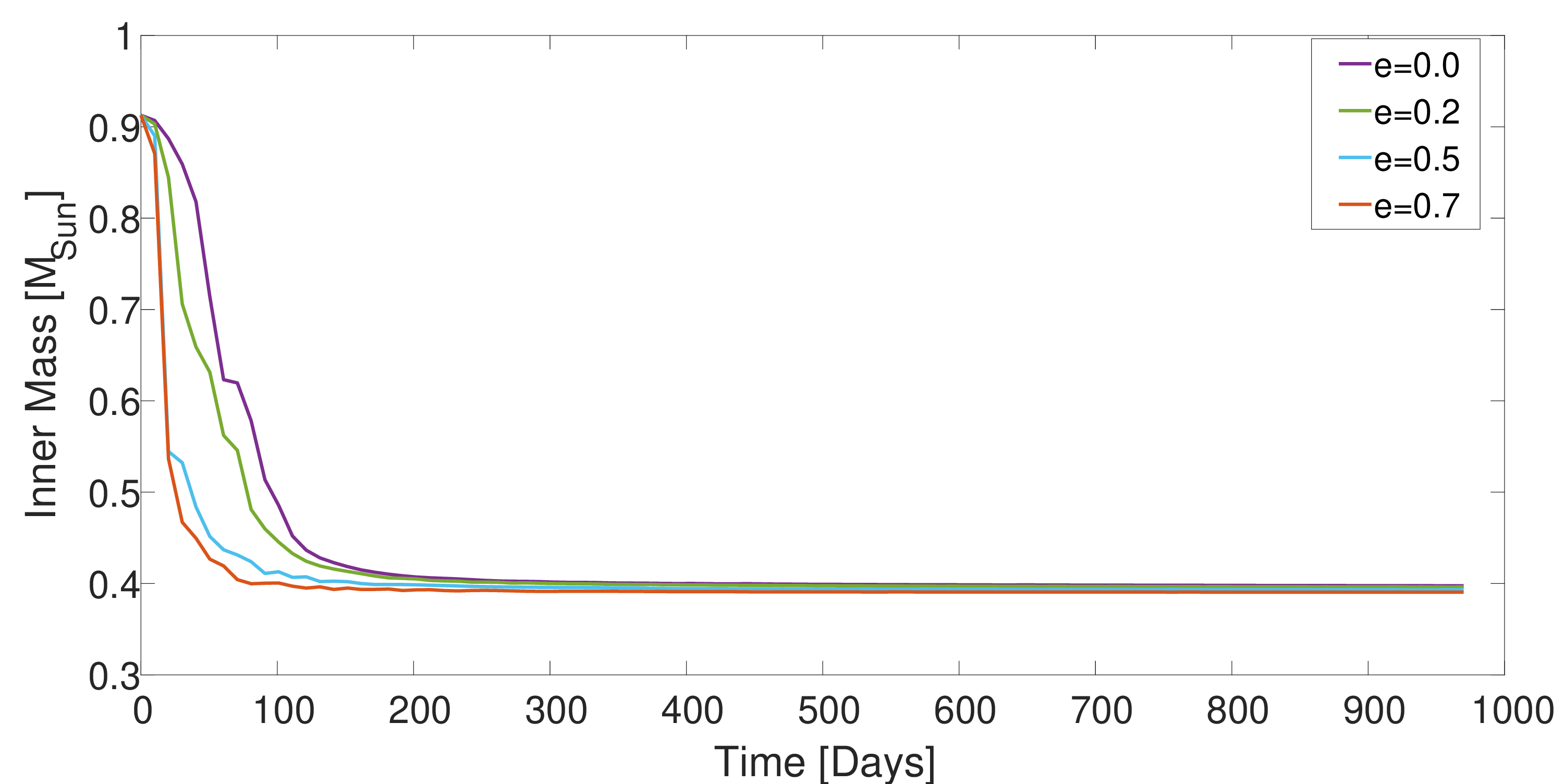}
    \includegraphics[width=\linewidth,clip]{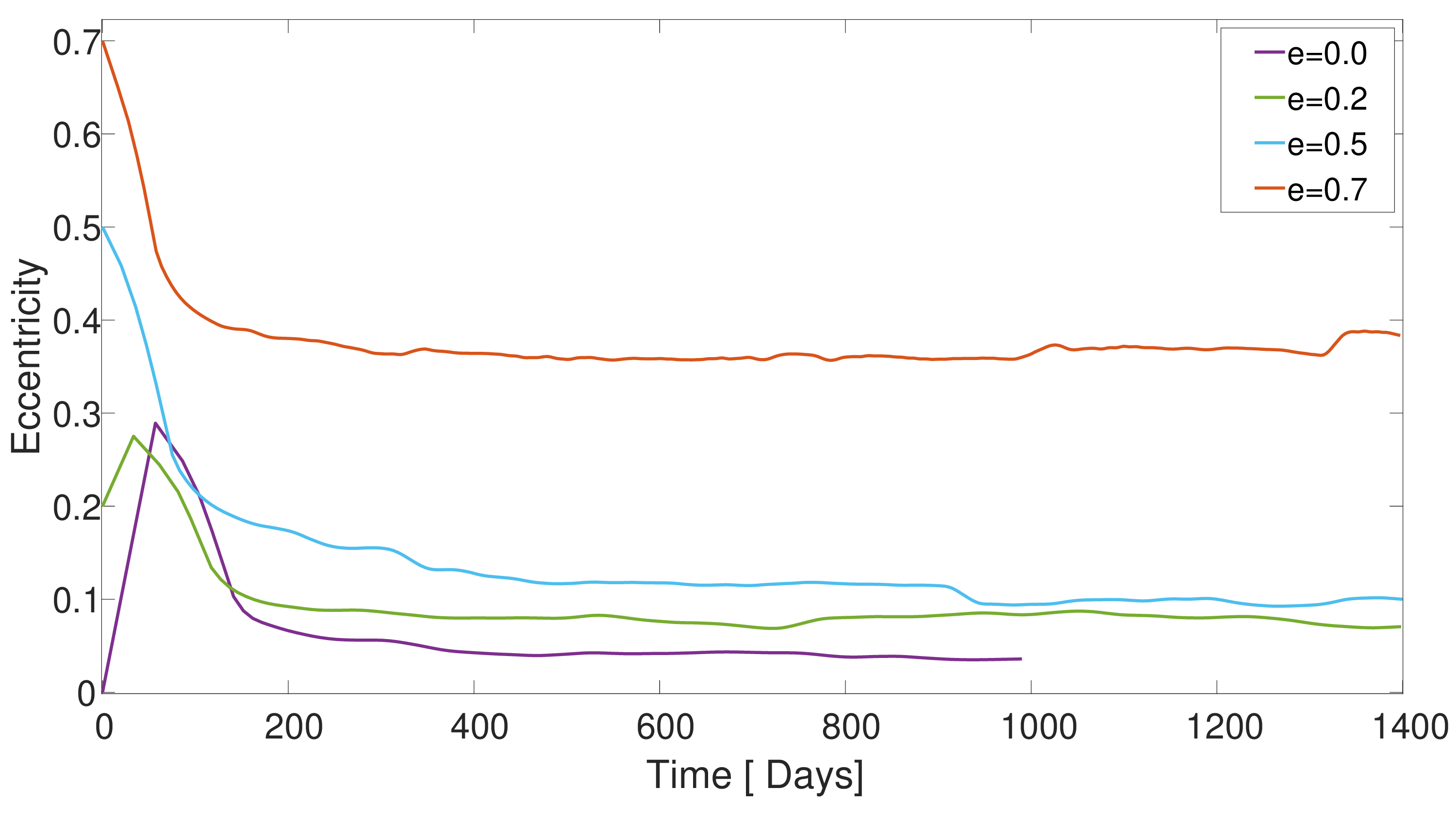}
    \includegraphics[width=\linewidth,clip]{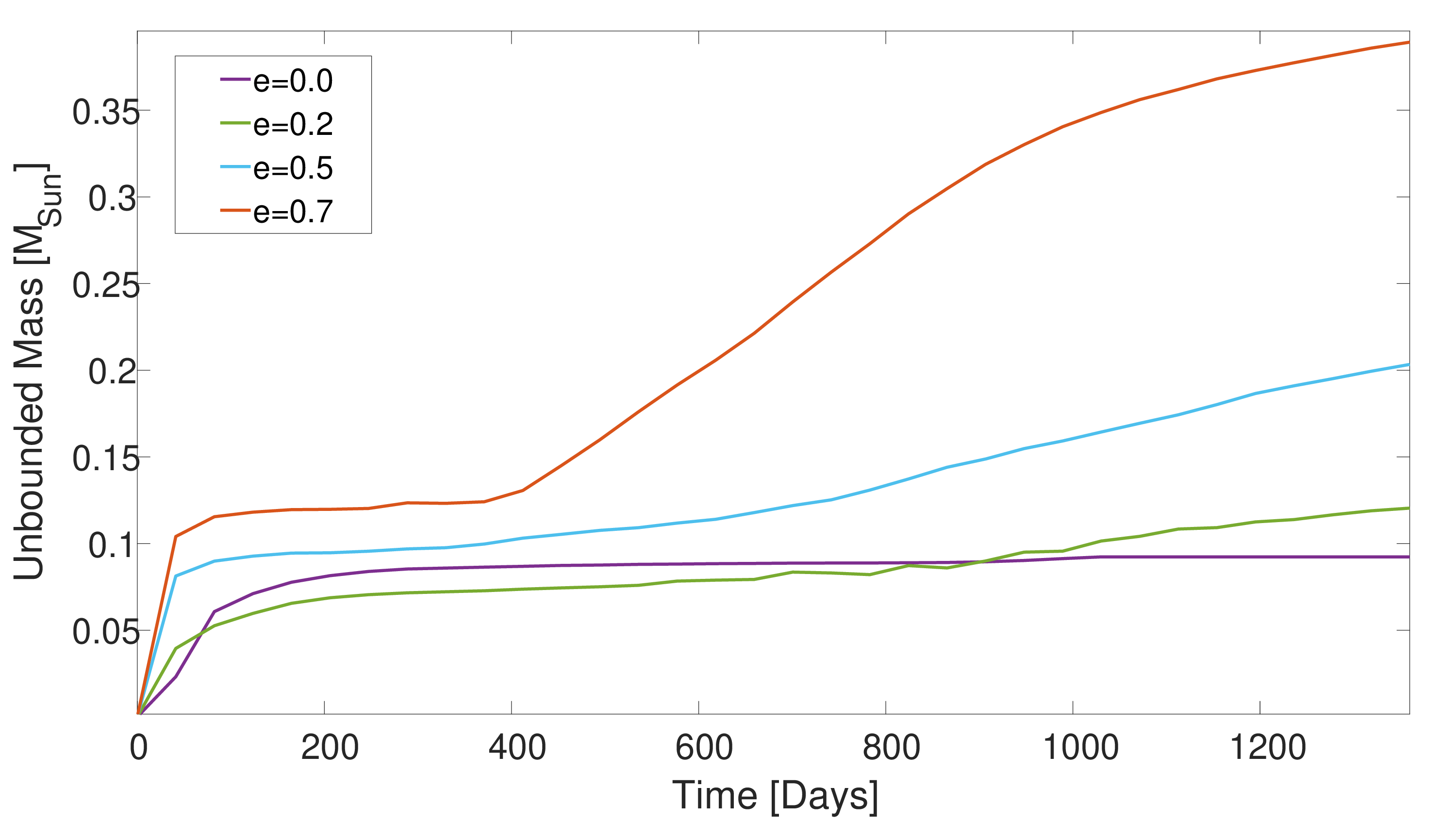}
    \caption{Upper panel: Mass of the giant (gas + core) located in the region between the two cores throughout the simulations with $1M_\odot$ giant and $0.6M_\odot$ companion, initialized with same distance (same apocenter distance) and different eccentricities.  Middle panel: quasi eccentricities.
    Lower panel: Calculated mass loss.}
    \label{fig:106a-innermass-eccentricity-massloss}
\end{figure}

Fig. \ref{fig:106a-orbit} and  \ref{fig:106a-separation} compare the separation between the primary core and the companion during simulations 1R06-0, 1R06A2-1R06A7 (see Tab. \ref{configurations-apo}), with varying eccentricities between 0 to 0.7. Since all of these systems were initialized with the giant already filling its Roche-lobe, and had their pericenter deeply inside the envelope of the primary, the orbit could not completely circularize. This can be seen in Fig. \ref{fig:106a-innermass-eccentricity-massloss}, which shows that the final eccentricities extends up to $e_f=0.4$, with smaller final separation for smaller initial pericenter. As a consequence of retaining such high eccentricities, the amounts of unbounded mass were higher, in comparison to simulations 1R06P2-1R06P95. Such results indicate the that amount of mass loss during the CEE relies strongly on the location of the pericenter.

In case of systems with same apocenter, an initially smaller eccentricity, corresponds to larger angular momenta $\left(\propto \sqrt{a_i(1-e_i^2)}\propto \sqrt{1-e_i}\right)$ and larger orbital energy $\left(\propto -a_i^{-1} \propto -(1+e_i) \right)$; thus, the companion efficiently unbinds the upper layers of the giant prior to stabilization. However, as the transferred energies and angular momenta go mostly to the upper layers, where the escape velocities are low, the amount of mass loss over the entire CE is larger for systems with initially smaller pericenter (higher eccentricities). Figure \ref{fig:106a-innermass-eccentricity-massloss} shows the calculated unbound mass throughout these simulations, showing that the ejection continue (mostly at pericenter passages) after the orbit has almost stabilized at fixed semimajor axis (same figure, upper panel) and fixed eccentricity (middle panel).
Nonetheless, in addition to the computed mass loss defined in Sec. \ref{massloss}, even after the region below the pericenter distance from the core has been completely diluted, further interactions at the apocenter lead to the increase in the ejected mass, that should further continue on a longer timescale, when interacting with the remaining bound mass. However, due to computational limitations we did not continue our simulations to measure this effect completely.

Simulations 8R2A0-8R2A7 began with initial apocenter more than twice the radius of the primary, where only the pericenter of the last 2 is initially inside the envelope, but with most of the first orbit outside the envelope (since $a_i > R_1$). All these simulations result with a core merger, and even in this case, the orbit could not circularize before the companion has reached the core. Since there was no synchronization, the amount of mass loss up to the merger point increases with the initial angular momenta and orbital energy of the system. Therefore, in these cases, the amount of mass loss increases for lower initial eccentricities. 
We can conclude, that the amount of mass loss is larger for larger initial angular momenta and orbital energy, unless the semi-major axis is initially located inside the envelope, leading to a much stronger unbinding, with smaller final separations.

\begin{figure}
    \centering
    \includegraphics[width=\linewidth,clip]{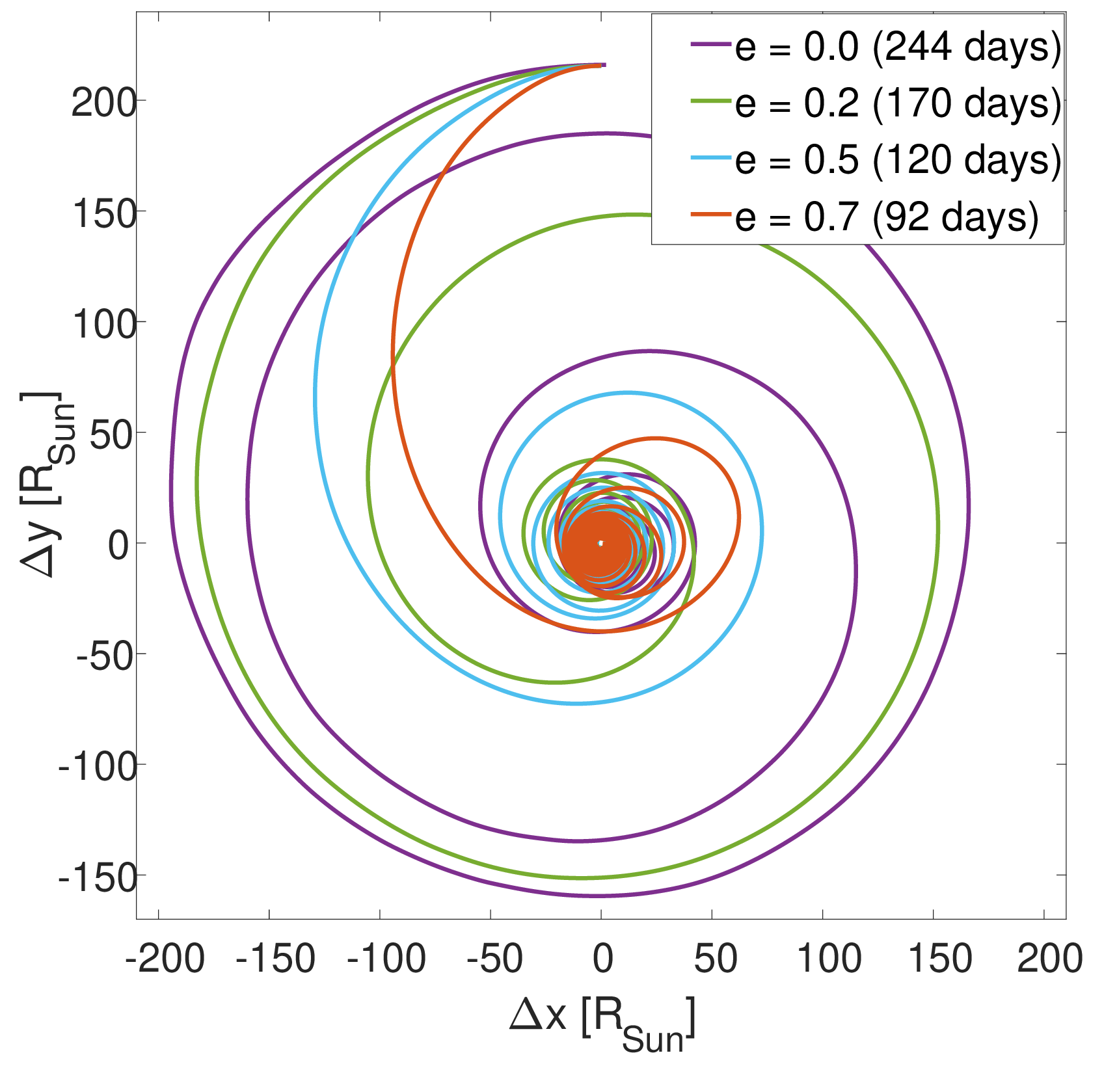}
    \caption{The orbit (distances between the two cores) of simulations with a $8M_\odot$ giant and a $2M_\odot$ companion, initialized with same distance (same apocenter distance) and different eccentricities}
    \label{fig:82a-orbit}
\end{figure}

Fig. \ref{fig:82a-orbit} shows a comparison of the orbital separations evolution of a different mass ratio, and initial apocenter more than twice the radius of the primary. Even in this case, the orbit  could not circularize before the companion has reached the core.

\begin{figure}
    \centering
    \includegraphics[width=\linewidth,clip]{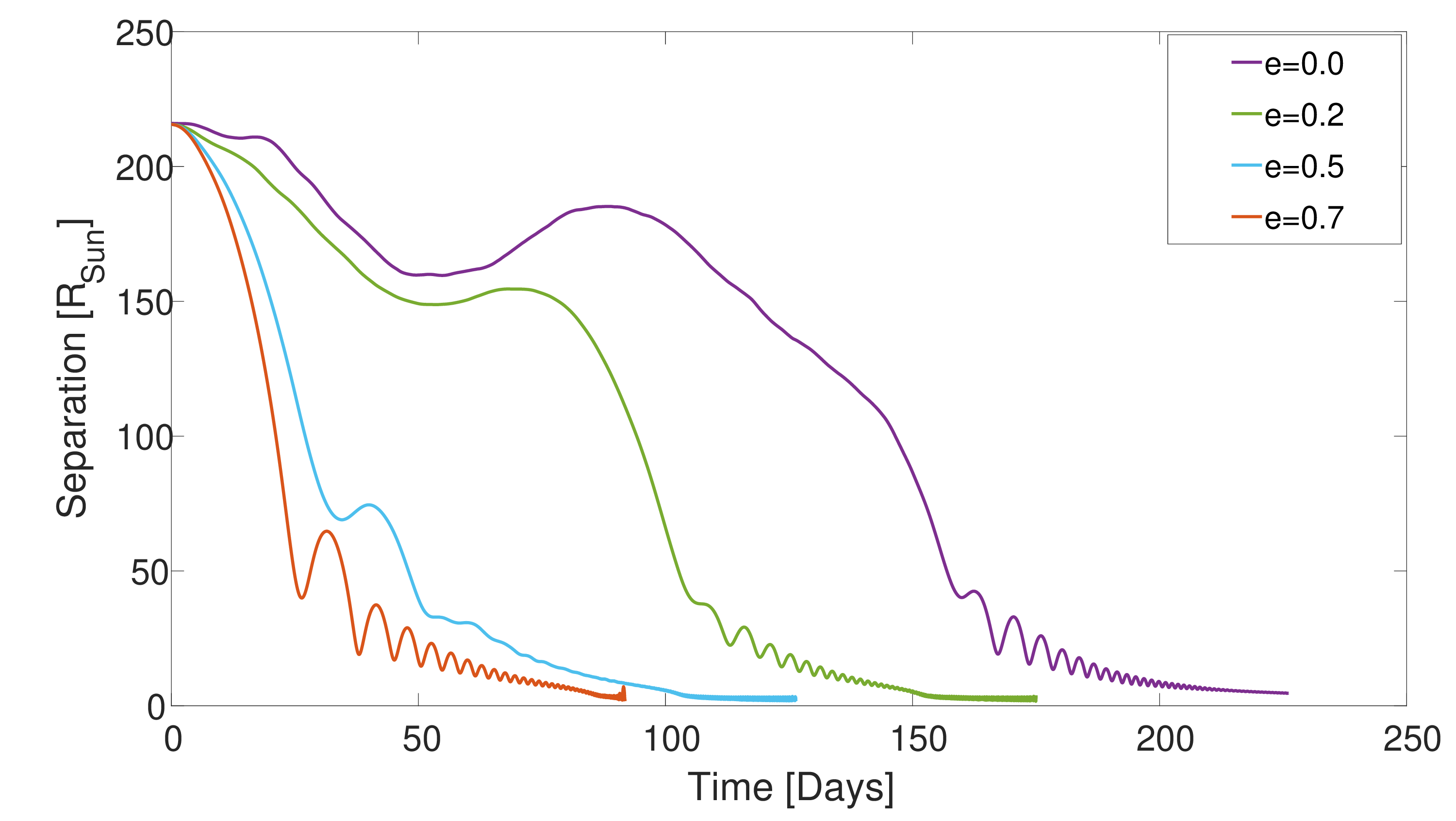}
    \includegraphics[width=\linewidth,clip]{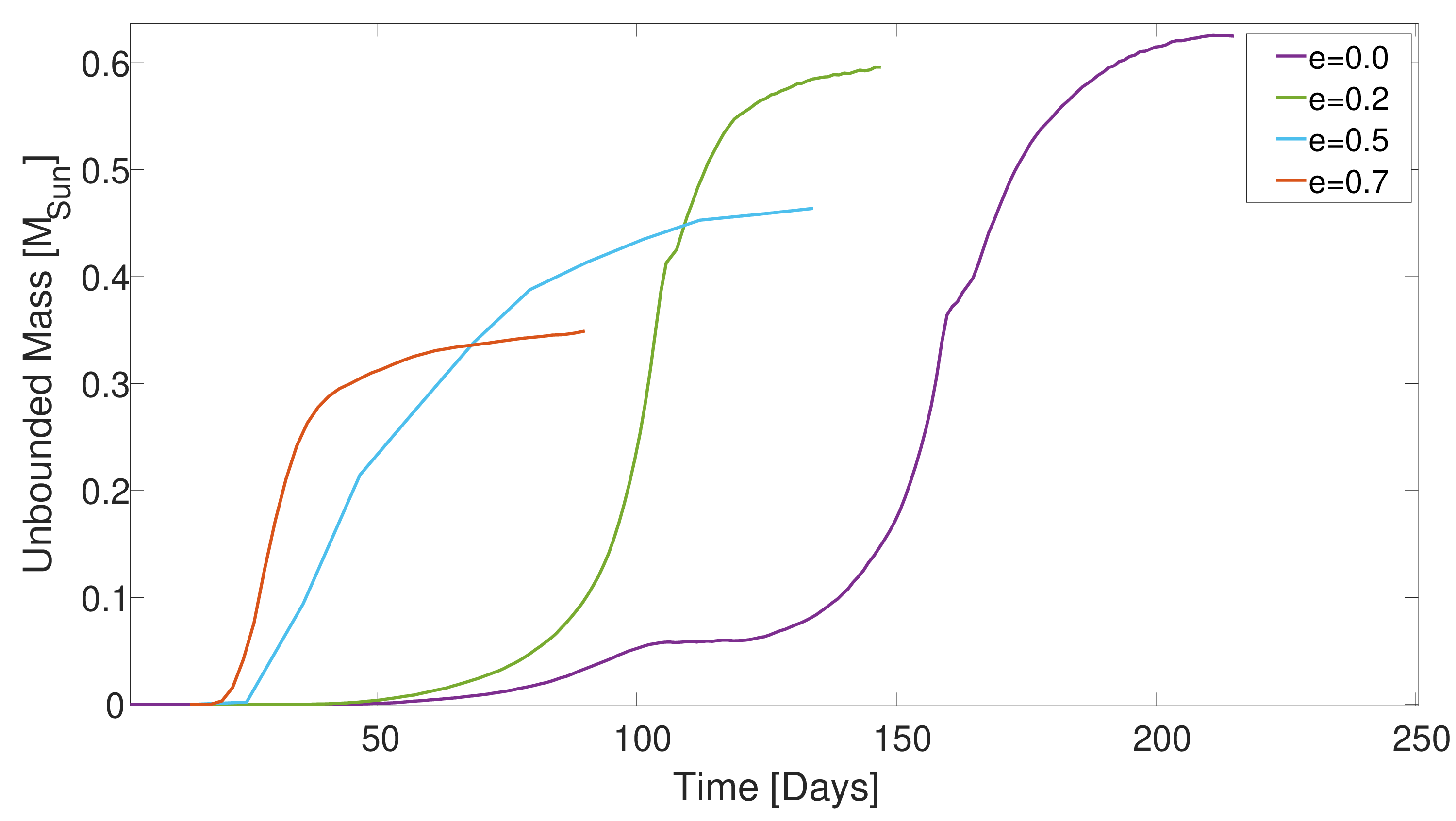}
    \caption{Separation between the primary's core and the companion of systems with an $8M_\odot$ giant and a $2M_\odot$ companion, initialized with same distance (same apocenter distance) and different eccentricities. Lower panel: calculated unbounded mass for same simulations.}
    \label{fig:82a-separation}
\end{figure}

\subsection{Eccentric common envelope evolution with an AGB primary star}
\begin{figure}
    \centering
    \includegraphics[width=\linewidth,clip]{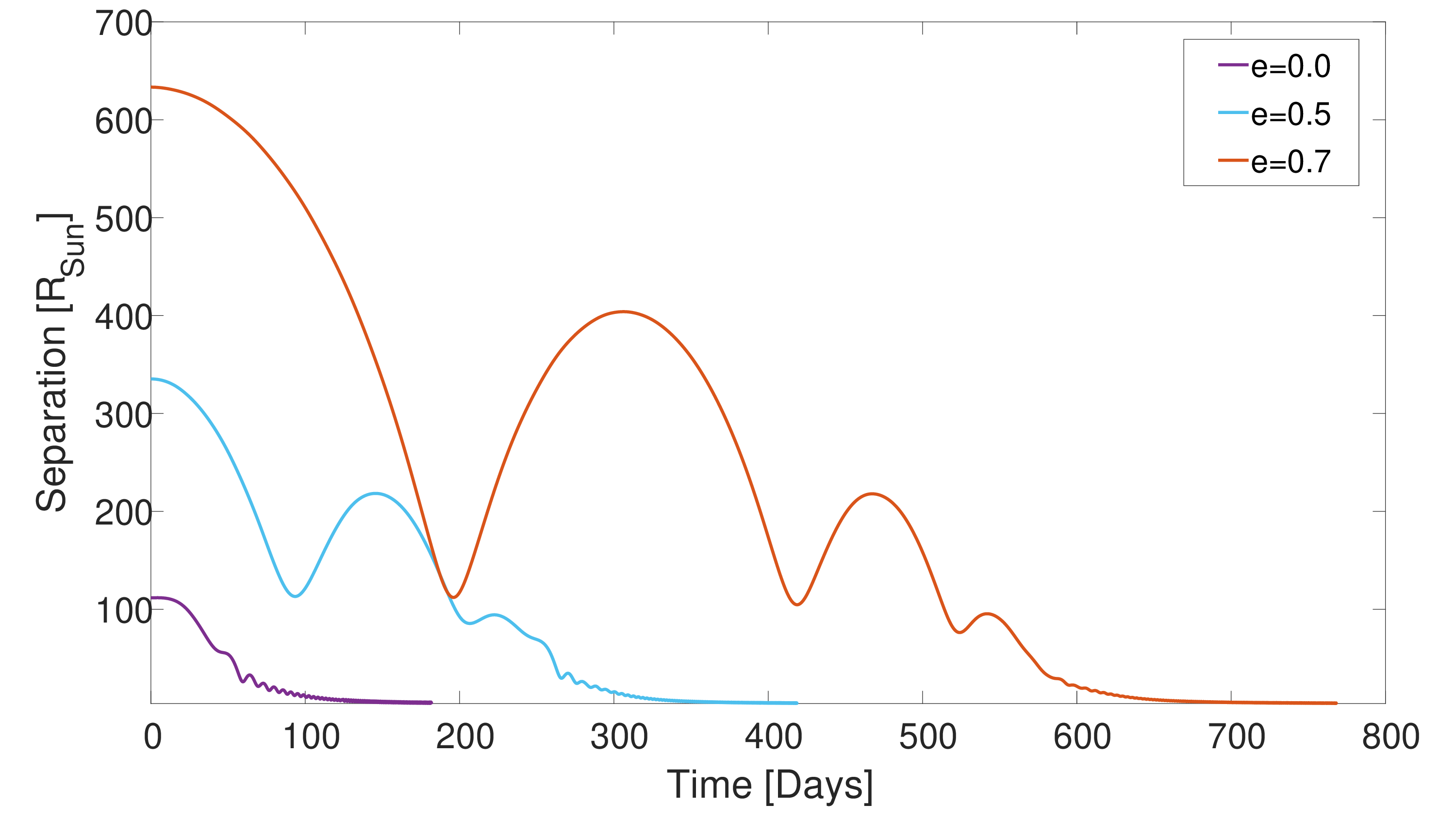}
    \includegraphics[width=\linewidth,clip]{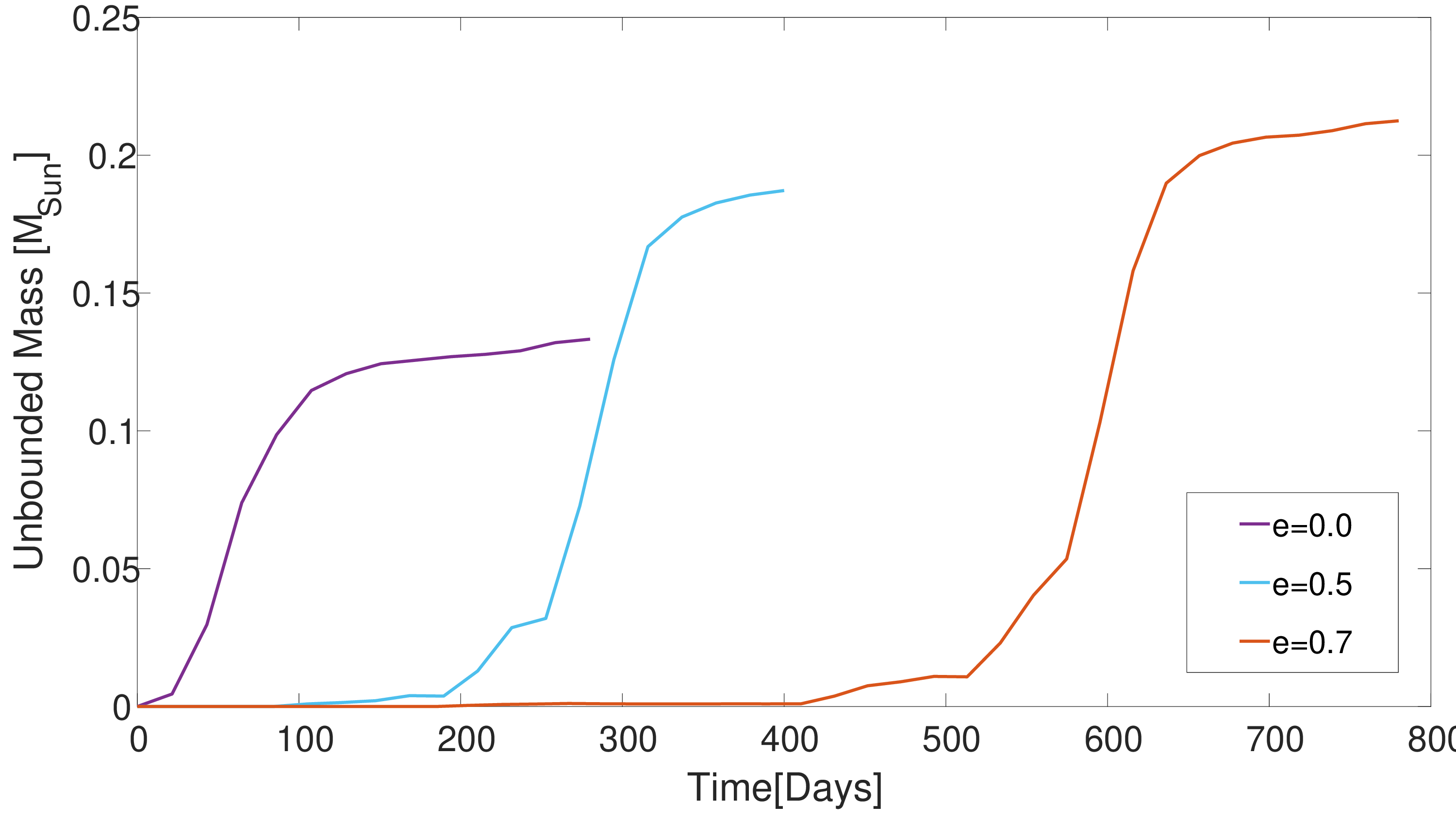}
    \caption{The separation between the primary's core and the companion of simulations with a $4M_\odot$ star on the Asymptotic giant branch and a $0.6M_\odot$ companion, initialized with different eccentricities, same pericenter distances and different initial distances (corresponding to their apocenters).}
    \label{fig:4a06p-separation}
\end{figure}

In simulations 4A06P0-4A06P7, we tested another scenario of a system with a $4M_\odot$ giant and a $0.6M_\odot$ companion, but now with the primary already at its AGB phase at the CE onset. We evolved the giant to $110R_\odot$, with a core mass of $M_{1,\text{core}}\approx0.698M_\odot$, and a fully convective envelope, when the giant is still stable during a dynamical timescale (see Fig. \ref{fig:4evolution}).  Due to its extended low-density envelope, the binding energy of an AGB star is much lower in comparison with the one at its RG stage. Consequently, the system can synchronize prior to the termination of the CEE, even in case of a core merger. In Fig. \ref{fig:4a06p-separation} we present the evolution of the separation between the two cores, and the mass loss. We can see that even though all three simulations ended with a core merger, all have a long self-regulating stage, when the spiral-in slows down significantly and the mass ejection terminates. As in all cases of primaries with RGs, one can identify the correlation between the mass ejection and the initial orbital eccentricity of the binary system, and despite the lower binding energy (which is still relatively high), all models ejected less than 10 percent of the envelope before the merger.
We note that the models of the AGB and RG differs in their density profiles, their cores, both in mass and radius, which affect their smoothing length profiles, and the initial energies and angular momenta. Therefore, a comparison between these cases is not in the scope of this work.

The ejected masses now show similar correlation with the initial eccentricities as in the simulations with the $8M_\odot$ giant, but the same claim about the ejected massed after releasing the energy and angular momenta of the binary at the merger, all asymmetries we find might disappear.

\begin{table*}
\begin{tabular}{|c|c|c|c|c|c|c|c|c|c|c|c|c|}
\hline 
 Sim & $M_\text{1,CEO}$ &$M_\text{2}$ &$M_{1,\text{core}}$ & $R_\text{1}$ &$r_{\text{i}}^{\text{a}}$ &$a_{\text{i}}$& $e_{\text{i}}$ & $R_{1,\text{RL}}$ & $r_i^p$ & $e_{\text{f}}$ & $M_{\text{unbound}}$ & $\frac{M_{\text{unbound}}}{M_\text{envelope}}$
\\ &(${\rm M_{\odot}}$)&(${\rm M_{\odot}}$) & $({\rm M_{\odot})}$ &(${\rm R_{\odot})}$&$({\rm R_{\odot})}$ &$({\rm R_{\odot})}$&&$({\rm R_{\odot})}$&(${\rm M_{\odot}}$)  & &$(M_\odot)$ & \tabularnewline
\hline 
\hline 
\hline 
1R06-0 & 0.913 & 0.6 & 0.388 &$83$ & 83 & 83 & 0.0 & 34 & 83 & 0.035 & 0.092 & 0.175 \tabularnewline
1R06P2 & 0.913 & 0.6 & 0.388 &$83$ & 124.5 & 103.8 & 0.2 & 52 & 83 & 0.038 & 0.1 & 0.19 \tabularnewline
1R06P5 & 0.913 & 0.6 & 0.388 &$83$ & 249 & 166 & 0.5 & 103 & 83 & 0.09 & 0.11 & 0.21\tabularnewline
1R06P7 & 0.913 & 0.6 & 0.388 &$83$ & 470.3 & 276.7 & 0.7 & 195& 83 & 0.11 & 0.12  & 0.229 \tabularnewline
1R06P9 & 0.913 & 0.6 & 0.388 &$83$ & 1577 & 830 & 0.9 & 668 & 83 & 0.14 & 0.146 & 0.279 \tabularnewline
1R06P95 & 0.913 & 0.6 & 0.392 &$83$ & 3237 & 1660 & 0.95 & 1372 & 83 & 0.18 & 0.136 & 0.261 \tabularnewline
\hline 
1R06A2 &0.913 & 0.6 & 0.388 &$83$ & 83 & 69 & 0.2 & 34 & 55.2 & 0.1 & 0.12 & 0.229 \tabularnewline
1R06A5 &0.913 & 0.6 & 0.388 &$83$ & 83 & 55.3 & 0.5 & 34 & 27.65 & 0.15 & 0.2 & 0.381 \tabularnewline
1R06A7 &0.913 & 0.6 & 0.388 &$83$ & 83 & 49 & 0.7 & 34 & 14.7 & 0.4 & 0.4 & 0.762\tabularnewline
\hline 
\hline 
8R2G-0 & 7.986 & 2 & 1.03 &110 & 115 & 115 & 0.0 & 58 & 115 & - & 0.4 & 0.058 \tabularnewline
8R2G2 & 7.986 & 2 &1.03 & 110 & 172.5 & 144 & 0.2 & 86 & 115 & - & 0.4 & 0.058 \tabularnewline
8R2G5 & 7.986 & 2 &1.03 & 110 & 345 & 230 & 0.5 & 173 & 115 & - & 0.48 & 0.069\tabularnewline
8R2G7 & 7.986 & 2 &1.03 & 110 & 651.7 & 383 & 0.7 & 326& 115 & - & 0.72 & 0.104\tabularnewline
\hline 
8R2-0 & 7.986 & 2 &1.03 & 110 & 216 & $216$ & 0.0 & 108 & 216 &- & 0.62 & 0.089 \tabularnewline
8R2A2 & 7.986 & 2 &1.03 & 110 & 216 & 180 & 0.2 & 108 & 144 & - & 0.6 & 0.086 \tabularnewline
8R2A5 & 7.986 & 2 &1.03 & 110 & 216 & 144 & 0.5 & 108 & 72 & - & 0.46 & 0.066 \tabularnewline
8R2A7 & 7.986 & 2 &1.03 & 110 & 216 & 127 & 0.7 & 108 & 38.1 & - & 0.35 & 0.05 \tabularnewline
\hline 
\hline 
4A06-0 & 3.967 & 0.6 &0.698 & 112 & 115 & 115 & 0.0 & 62.7 & 115 & 0.1 & 0.14 & 0.043 \tabularnewline
4A06P5 & 3.967 & 0.6 & 0.698 &112 & 345 & 230 & 0.5 & 188 & 115 & 0.1 & 0.19 & 0.058 \tabularnewline
4A06P7 & 3.967 & 0.6 & 0.698 &112 & 651.7 & 383 & 0.7 & 355.3 & 115 & 0.1 & 0.21 & 0.064\tabularnewline
\hline 
\hline 
4R06-0 & 3.992 & 0.6 & 0.478&42 & 43 & 43 & 0.0 & 23.4  & 43 & - & 0.13 & 0.037 \tabularnewline
4R06P2 &  3.992 & 0.6 & 0.478&42 & 64.5 & 54 & 0.2 & 35.1 & 43 & - & 0.144 & 0.041 \tabularnewline
4R06P5 &  3.992 & 0.6 & 0.478&42 & 129 & 86 & 0.5 & 70.3 & 43 & - & 0.177 & 0.05 \tabularnewline
4R06P7 &  3.992 & 0.6 & 0.478&42 & 243.7 & 143.3 & 0.7 & 133 & 43 & -  & 0.18 & 0.051 \tabularnewline
4R06P9 &  3.992 & 0.6 & 0.478&42 & 817 & 430 & 0.9 & 445.4 & 43 & - & 0.24 & 0.068  \tabularnewline
4R06P95 &  3.992 & 0.6 & 0.478&42 & 1677 & 860 & 0.95 & 914.2 & 43 & - & 0.22  & 0.063 \tabularnewline
\hline 
\end{tabular}\caption{\label{results-table} Initial configuration of the simulated systems at the onset of the CE, and final values. $M_\text{1,CEO}$, $R_1$ and $M_{1,\text{core}}$ are the mass, radius and core mass of the primary at the beginning of the simulation, $M_\text{2}$ is the mass of the secondary, $r_{\text{i}}^{\text{a}}$, $r_{\text{i}}^{\text{a}}$ are the initial apoapsis and periapstron distances, $a_{\text{i}}$ is the initial semi-major axis of the binary system, $e_{\text{i}}$, $e_{\text{f}}$ are the initial and final eccentricities, $R_{1,\text{RL}}$ is the initial Roche-Lobe radius of the giant, and $M_{\text{unbound}}$ is the calculated unbounded mass. }
\end{table*}

\section{Discussion}
\label{sec:Discussion}

\subsection{Comparison with eccentricities of observed post common envelope binaries}
\label{eccentricPCE}
In most cases, the orbital eccentricity of observed post-CE binary systems that contain either SdB, SdO or WD components, was not measured (and in some cases the orbital are just assumed to be circular). However, when available, the data on the eccentricities of such systems show eccentricities comparable to those we find in our simulations in the range of a few 0.001 to 0.18 (a few with an upper boundary of up to 0.64) and in particular in cases of masses comparable to our cases, with the vast majority having $e<0.03$ \citep{1999Delfosse-eccentric-short-Mdward-WD,2005Edelmann-eccentric-short-sdBs, 2015Kawka-closesdBseccentric, 2021KruckowPotentialPCE}.

\subsection{Mass loss during the common envelope phase}
\label{massloss}
At the end of the common envelope phase, the envelope is believed to be completely ejected \citep{Iva+13}. However, to our knowledge, most of the previous numerical simulations ended with most of the envelope still bound \citep{Pass+12,Ric+12,Iva+13}. A few works which assumed a complete local thermalization of the recombination energy during the spiral-in gave rise to larger or complete unbinding \citep{Nan+15,2020SandAGBCEEjectionWithRecombinationOPAL,2020KramerSdBCEEjectionRecombinationOPAL} ,although the definition of unbinding also differs between studies in which the thermal energy is included in the definition, i.e. assuming it effectively translates to kinetic energy, or not; see more below. Nevertheless, consistent modelling of the role of recomobination is difficult, and its imprtance and overall role in envelope ejection is still debated \citep{Han+94,Har+98,2002HanIonization,Sok+03,Iva+15,sabachconvection,2018AldanaRecombination,Reichardt2020,2020SandAGBCEEjectionWithRecombinationOPAL}. Note, that the definition of bound material includes the thermal energies in some cases, but not in others. This could pose a major problem, since remaining bound mass might infall on longer timescales and eventually lead to the merger of any post CE binaries, in contrast with observations of such binaries \citep{Iva+13,2005Edelmann-eccentric-short-sdBs,Siess_2014,2015Kupfershortpeiod,2015Kawka-closesdBseccentric,2017A&A-vos-sdbMS,Ratzloff2020sdoandwd}. These results suggest the potential importance of some processes which were not considered in those simulations as mentioned in the introduction.

In this paper, we do not aim to solve the mass loss problem, but we do want to measure its connection to the orbital eccentricity.
We calculate the amount of mass lost at any point of the simulation by summing the mass of all SPH particles with a positive energy, define  as:
\begin{equation}
    E_i=E_{\text{kin},i}-E_{\text{pot},i}
\end{equation}
where $E_{\text{kin}.i}$ and $E_{\text{pot},i}$ are the kinetic, thermal and potential energies of the particle with index $i$. The potential energy is calculated as follows:
\begin{equation}
    E_{\text{pot},i}=\sum_{\text{core},\text{comp},j\ne i}\frac{Gm_im_j}{r_{i,j}}
\end{equation}
where $m_i$ is the mass of particle i and $r_{i,j}$ is the distance between particles i and j. One can see that the potential energy is calculated with respect to all simulated particles, both bound and unbound. While a true condition for being bound to a system is not affected by external potentials. As a consequence, when the unbound mass fraction is higher, the number of unbound particles which affect our calculations is larger, and thus the uncertainty of the actual mass loss is increased.
In order to consider and compare the ejected massed, we measure the mass loss using this condition, during the last stages of each simulation, and take the value once the system is stabilized (no significant evolution of the binary orbit).

\subsection{The time scales for envelope ejection}
\label{timescales}
\cite{soker-eccentricitylossinagb} showed that the orbital eccentricity of a binary system inside a CE increases significantly with the mass loss rate. However, as was showed in our results, the circularization of the orbit by tidal forces is very strong, leaving our simulated system with an initial pericenter outside the envelope, with a relatively small final eccentricity even for the largest initial eccentricity. Therefore, an additional efficient mass ejection during the spiral-in should lead to larger final eccentricities at its termination. Nevertheless, interactions of the post-CE system with the surrounding gas may cause an additional eccentricity amplification at apocenter interactions \citep{1991ArtymowiczEccentricityGrow, 2011MNRASKashiCircumbinaryDisk, 2018MNRAS-sdBSampleEccentric-vos}. Therefore, if most of the mass ejection is due to the potential energy dissipation during the spiral-in (with some additional mass loss occurring later-on, e.g. as in the case of dust driven winds), we expect to find most post-CE binaries located inside planetary nebulae to have low eccentricities, if any, whereas post-CE binaries with a complete unbound nebulae to have larger eccentricities. If additional processes cause a larger mass loss rate during the spiral-in, we should expect the majority of the post-CE binaries to have even larger eccentricities, possibly inconsistent with current post-CE binaries.

\subsection{Implications for compact object mergers: supernovae, gamma-ray bursts and gravitational-wave sources}
Our findings suggest that post-CE binaries could have non-negligible remnant eccentricities following the CE phase. Many of the mergers of compact object binaries are thought to occur following a CE phase which shrinks the binary orbits, which is then followed by a typically much slower inspiral due to gravitational-wave (GW) emission. The latter, GW inspiral is sensitive to the initial eccentricity of the binary, where, for a given SMA- eccentric binaries merge faster than more circular ones \cite{Peters1964}. In other words, remnant eccentricities could potentially affect delay time distribution of mergers, and thereby affect the delay time distributions of type Ia supernovae that result from mergers of WD, short Gamma-ray burst resulting from neutron stars (and possibly black hole) mergers and GW sources from the mergers of WD, NSs and/or BHs. Current population synthesis models inferring the delay-time distributions of such transient explosive events typically assume CEE always leads to final circular orbits. Our results suggest that such assumptions might be erroneous and should potentially be corrected. The eccentricities we find are typically not extremely high, and therefore we expect the effects to change the inspiral times by a few up to $\sim20\%$, however it could differ by as much as a factor of two for binaries which initial pericenters are embedded deep inside the giant envelope.
 
 In principle, remnant eccentricities might also lead to eccentric GW sources, but our derived eccentricities are likely too low. Although a GW inspiral circularizes a binary, an initially highly eccentric binary might retain some eccentricity up to its merger, potentially observable by GW detectors. However such cases typically require very close and very eccentric orbits ($1-e\ll1)$, higher than expected from our models.    
 
\section{Summary}
\label{sec:Summary}
In this work, we used hydro-dynamical simulations of the common envelope evolution to study the effect of different initial eccentricities on the evolution throughout this stage. 
We used the \texttt{AMUSE} framework \citep{AMUSE-Book} to combine the stellar evolution of the giants prior to the common envelope, simulated with \texttt{MESA} \citep{mesa-article}, and the hydrodynamical evolution of the binary system, simulated with the SPH code \texttt{GADGET2} \citep{gadget2}.
We simulated 26 systems with different mass ratio of the binary components, and a large variety of different initial eccentricities, up to 0.95, which initiated with a very large apocenter, required even higher computational resources than those usually used for CE simulations.
We find that CEE of initially eccentric binaries give rise to remnant post-CE binaries with non-negligible eccentricities, with up to 0.18 (0.4) for post-CE binaries with initial peri-centers outside (inside) the giant star. The post-CE eccentricities appear to be correlated with the initial eccentricities.
We note that population synthesis studies typically assume post-CE binaries have circular orbits. Our results suggest that such modelling might be inconsistent both with our models and observed systems, and that accounting for remnant eccentricities could affect the typical timescales and delay time distributions for the mergers of compact binaries and their resulting explosive electromagnetic and GW transients, such as supernovae gamma-ray bursts and GW-sources.     

Finally, we find that the amount of mass loss during the evolution and the structure of the post-CE nebulae strongly depend on the location of the pericenter and the initial orbital energy. 
All simulations with the initial pericenter located outside the envelope show relatively low final eccentricities, in agreement with current observed post-CE systems. This suggests that the still bound envelope that remain at the end of the CEE, is eventually ejected on much longer timescales than the CEE inspiral, as suggested by \cite{Michely+19, 2019igoshev}.

\section*{Acknowledgements}
This work was done during the world-wide Covid-19 pandemic.
We thank Noam Soker, Alexey Bobrick and Aldana Grichener for helpful discussions. HBP acknowledge support for this project from the European Union's Horizon 2020 research and innovation program under grant agreement No 865932-ERC-SNeX. 
We used the following codes in the simulations, analysis and visualizations presented in this paper: \texttt{AMUSE} \citep{amusearticle,AMUSE-Book}, \texttt{MESA} (version 2208) \citep{mesa-article}, \texttt{GADGET2} \citep{gadget2}, \texttt{HUAYNO} \citep{2012NewA...17..711P}, \texttt{matplotlib} \citep{matplotlib}, \texttt{pynbody} \citep{pynbody}, \texttt{NumPy} \citep{numPy}. All models were running on  the Astric computer cluster of the Israeli I-CORE center.

\section*{Data Availability}
All data underlying this research is available upon reasonable request to the corresponding authors.



\bibliographystyle{mnras}


\bsp	
\label{lastpage}
\end{document}